\documentclass[twocolumn]{aa}  

\usepackage{color}
\usepackage{txfonts}
\usepackage[version=4]{mhchem}
\usepackage{placeins}

\begin{document}

\title{CO destruction in protoplanetary disk midplanes: inside versus outside the CO snow surface}
\author{Arthur D. Bosman \inst{1}, Catherine Walsh \inst{2}, Ewine F. van Dishoeck \inst{1,3} }
\institute{Leiden Observatory, Leiden University, PO Box 9513, 2300 RA Leiden, The Netherlands\\ \email{bosman@strw.leidenuniv.nl},
\and
School of Physics and Astronomy, University of Leeds, Leeds, LS2 9JT, UK
\and
Max-Planck-Insitut f\"{u}r Extraterrestrische Physik, Gie{\ss}enbachstrasse 1, 85748 Garching, Germany}
\abstract
{The total gas mass is one of the most fundamental properties of disks
around young stars, because it controls their evolution and their potential to form planets. 
To measure disk gas masses, CO has long been thought to be the best
tracer as it is readily detected at (sub)mm wavelengths in many
disks. However, inferred gas masses from CO in recent ALMA
observations of large samples of disks in the 1--5 Myr age range seem
inconsistent with their inferred dust masses. The derived gas-to-dust
mass ratios from CO are between one and two orders of magnitude lower than the ISM
value of $\sim$ 100 even if photodissociation and freeze-out are
included. In contrast, \textit{Herschel} measurements of hydrogen deuteride line
emission of a few disks imply gas masses in line with gas-to-dust
mass ratios of 100. This suggests that at least one additional mechanism is
removing \ce{CO} from the gas-phase. }
{Here we test the suggestion that
the bulk of the CO is chemically processed and that the carbon is
sequestered into less volatile species such as \ce{CO2}, \ce{CH3OH}
and \ce{CH4} in the dense, shielded midplane regions of the disk. This
study therefore also addresses the carbon reservoir of the material
which ultimately becomes incorporated into planetesimals.}
{Using our gas-grain chemical code we performed a parameter exploration
and follow the CO abundance evolution over a range of conditions
representative of shielded disk midplanes. }
{Consistent with previous studies, we find
that no chemical processing of CO takes place on 1--3 Myr timescales
for low cosmic-ray ionisation rates, $< 5\times 10^{-18}$ s$^{-1}$. Assuming
an ionisation rate of $10^{-17}$ s$^{-1}$, more than 90\% of the CO is converted into other species, but only in the cold parts of the disk below 30 K.
This order of magnitude destruction of CO is robust against the choice
of grain-surface reaction rate parameters, such as the tunnelling efficiency and
diffusion barrier height, for temperatures between 20 and 30 K.  Below
20 K there is a strong dependence on the assumed efficiency of H
tunnelling.}
{The low temperatures needed for CO chemical processing indicate that
the exact disk temperature structure is important, with warm disks
around luminous Herbig stars expected to have little to no \ce{CO}
conversion. In contrast, for cold disks around sun-like T Tauri stars, a
large fraction of the emitting CO layer is affected unless the disks are young ($< 1$ Myr). This can lead to
inferred gas masses that are up to two orders of magnitude too
low. Moreover, unless CO is locked up early in large grains, the
volatile carbon composition of the icy pebbles and planetesimals forming in the 
midplane and drifting to the inner disk will be dominated by CH$_3$OH,
CO$_2$ and/or hydrocarbons.}

\keywords{protoplanetary disks -- astrochemistry -- molecular processes -- ISM: molecules}

\authorrunning{A. D. Bosman et al.}
\titlerunning{CO destruction in protoplanetary disk midplanes}
\maketitle

\section{Introduction}
The total gas mass is one of the most fundamental parameters that influences
protoplanetary disk evolution and planet formation.  Interactions of
the gas and dust set the efficiency of grain-growth and planetesimal
formation \citep[e.g. ][]{Weidenschilling1977, Brauer2008, Birnstiel2010,Johansen2014}, while interactions of planets
with the gaseous disk leads to migration of the planet and gap
formation \citep[see, e.g.][for reviews]{Kley2012,Baruteau2014}. Significant amounts of
gas are needed to make giant Jovian-type planets. All of these
processes depend sensitively on either the total amount of gas or the
ratio of the gas and dust mass. Dust masses can be estimated from the
continuum flux of the disk, which is readily detectable at
sub-millimeter (mm) wavelengths. However, the main gaseous component
\ce{H2} does not have any strong emission lines that can trace the
bulk of the disk mass, so that other tracers need to be used.
Emission from the \ce{CO} molecule and its isotopologues is commonly
used as a mass tracer of molecular gas across astronomical
environments \citep[for reviews see,
  e.g. ][]{Dishoeck1987,Bolatto2013,Bergin2017}. \ce{CO} is resistant
to photodissociation because it can self-shield against UV-photons and
is thus a molecule that can trace \ce{H2} in regions with low dust
shielding \citep{Dishoeck1988,Viala1988,Lee1996,Visser2009}. \ce{CO}
also has, in contrast with \ce{H2}, strong rotational lines, coming
from states that can be populated at 20 K, the freeze-out temperature
of \ce{CO}. In most astronomical environments, \ce{CO} is also
chemically stable due to the large binding energy of the C--O
bond. This chemical stability means that \ce{CO} is usually the second
most abundant gas-phase molecule and the main volatile carbon
reservoir in molecular astronomical environments. Thus, the recent
finding that CO emission from protoplanetary disks is very weak came
as a big surprise and implies that CO may be highly underabundant
\citep{Favre2013,Bruderer2012, Du2015, Kama2016,Ansdell2016}. Is CO
transformed to other species or are the majority of disks poor in gas
overall?

By extrapolating the chemical behaviour of \ce{CO} from large scale
astronomical environments to protoplanetary disks it was expected that
only two processes need to be accounted for in detail to determine the
gaseous \ce{CO} abundance throughout most of the disk:
photodissociation and freeze-out of \ce{CO}
\citep{Dutrey1997,vanZadelhoff2001,vanZadelhoff2003}. This was the
outset for the results reported by \cite{Williams2014} who computed a
suite of disk models with parametrised chemical and temperature
structures, to be used for the determination of disk gas masses from
the computed line emission of CO isotopologues. This method was
expanded by \cite{Miotello2014, Miotello2016} who calculated the
temperature, \ce{CO} abundance and excitation self-consistently using
the thermo-chemical code
DALI\footnote{\url{http://www.mpe.mpg.de/~facchini/DALI/}}
\citep{Bruderer2012, Bruderer2013}. \cite{Miotello2016} used a simple
gas-grain network that includes CO photodissociation, freeze-out and
grain-surface hydrogenation of simple species, but no full grain
surface chemistry. DALI also computes the full 2D dust and gas
temperature structure, important for determining the regions affected
by freeze-out and emergent line emission. Because emission from the
main \ce{CO} isotopologue \ce{^{12}C^{16}O} is often optically thick,
most observations target the rarer \ce{CO} isotopologues. These do not
necessarily follow the highly abundant \ce{^{12}C^{16}O} as
\ce{^{12}C^{16}O} can efficiently shield itself from photodissociating
UV radiation at lower \ce{H2} column densities compared with the less
abundant isotopologues. As such, the rarer isotopologues are
dissociated over a larger region of the disk, an effect known as
isotope-selective photodissociation \citep[see, for
  example][]{Visser2009}. The combined effects of the different
temperature structure and isotope-selective photodissociation change
the emission strengths of the CO isotopologues by up to an order of
magnitude compared with the predictions of \cite{Williams2014}.

When either of these model predictions including photodissociation and freeze-out are applied to ALMA
observations of large samples of disks, still low gas masses are determined:
inferred gas masses are close to, or lower than, the calculated dust
mass from the same observations instead of the expected 100:1 ratio
 \citep{Ansdell2016, Miotello2017,
  Pascucci2016, Long2017}. While it is possible that these disks are
indeed very gas depleted, independent determinations of the gas masses
such as from far-infrared \ce{HD} data \citep[see,
  e.g.][]{Bergin2013,McClure2016,Trapman2017} and mass accretion rates
\citep{Manara2016b} imply that the \ce{CO}/\ce{H2} abundance ratio is likely
much lower than expected, at least in the CO emitting part of the disk.

Multiple mechanisms have been proposed to explain this low \ce{CO} abundance, both chemical and physical. A physical argument for the low \ce{CO} abundances comes from the vertical mixing of the gas together with settling of dust. \cite{Kama2016} argued that the low CO abundance in the upper emitting layers of the outer disk can be explained by the constant vertical cycling of gaseous \ce{CO}. Every vertical cycle some \ce{CO} will freeze-out onto grains that have grown and settled below the \ce{CO} snow surface. These larger grains do not cycle back up again to the warmer regions where CO can be returned to the gas. They show that the \ce{CO} abundance can be significantly lowered over the disk lifetime. This mechanism also predicts a strong anti-correlation between age and measured \ce{CO} abundance. The mechanism can explain the destruction of \ce{CO} in the warm layers, such as reported by \cite{Schwarz2016} and at the same time explain the lower than expected \ce{H2O} abundances found in the outer disk of TW Hya and other disks by \cite{Hogerheijde2011} and \cite{Du2017}. However this mechanism cannot explain the low abundance of \ce{CO} inside of the \ce{CO} iceline, the radial location of the snow surface at the midplane, as inferred by \cite{Zhang2017} for TW Hya. 

Alternatively, there are various chemical mechanisms that destroy \ce{CO}, sometimes referred to as `chemical depletion'. Some of the proposed chemical pathways start with the destruction of gaseous \ce{CO} by \ce{He+}, leading to the formation and subsequent freeze-out of \ce{CH4} \citep{Aikawaflow1999, Eistrup2016} or, when computed at slightly higher temperatures, the gas-phase formation of \ce{C2H2} and subsequent freeze-out and further chemical alteration on the grain-surface \citep{Yu2016}. Another pathway to destroy \ce{CO} is the reaction with \ce{OH} to form \ce{CO2}, either in the gas-phase \citep{Aikawaflow1999}, or on the grain-surface \citep{Furuya2014, Reboussin2015, Drozdovskaya2016, Eistrup2016, Schwarz2018}. The formation of \ce{CO2} through the grain-surface route seems to be most efficient at temperatures around 25 K, just above the freeze-out temperature of \ce{CO}. A third pathway to destroy \ce{CO} is the hydrogenation of \ce{CO} on the dust grain-surface forming \ce{CH3OH} \citep{Cuppen2009, Yu2016, Eistrup2018}. All of these models start with a high abundance of CO and modify the abundance through chemical processes. Alternatively there models that do not have CO initially as they assume that, due to some reset process, the gas is fully ionised or atomic at the start of the calculation \citep{Eistrup2016,Molyarova2017}. Due to the high abundance of \ce{OH} during the transition of atomic to molecular gas, \ce{CO2} can be efficiently formed. At low temperatures (< 50 K) \ce{CO2} becomes the most abundant carbon bearing species.  

All of these \ce{CO} destruction processes are driven by dissociating
or ionising radiation, either UV photons, X-rays or cosmic-rays. In
regions where UV photons and X-rays are not able to penetrate,
cosmic-rays drive the chemistry, so that the chemical timescales of
\ce{CO} processing are strongly dependant on the cosmic-ray ionisation
rate \citep{Reboussin2015, Eistrup2018}. Indeed, \cite{Eistrup2016}
show that chemical evolution during the disk lifetime in the dense midplane 
is negligible if the only
source of ionisation is provided by the decay of radioactive
nuclides. In line with these results, \cite{Schwarz2018} find that even in the
warm molecular layers either a cosmic-ray ionisation rate of
$10^{-17}$ s$^{-1}$ or a strong X-ray field is needed to significantly
destroy \ce{CO}. High cosmic-ray ionisation rates are not expected if
the proto-stellar magnetic field is sufficiently strong to deflect
galactic cosmic-rays \citep{Cleeves2015}.

The goal of this paper is to study the chemical pathways that can
destroy \ce{CO} in those regions of the disk that are sufficiently
shielded from UV photons such as that near the disk midplane. The
effectiveness and timescale of \ce{CO} destruction pathways as
functions of temperature, density and cosmic-ray ionisation rate are
investigated for comparison with the increasing number of ALMA
  surveys of CO in disks in the 1-10 Myr age range. We also study the
effect of the assumed grain-surface chemistry parameters, in
particular the tunnelling barrier width and the diffusion-to-binding
energy ratio. To be able to do this study in an as general sense as
possible we do not restrict ourselves to any specific disk structure
but instead perform a parametric study of temperature, density and
cosmic-ray ionisation rate over a range representative of a
significant portion of the disk mass.

\section{Methods}
\subsection{Parameter space}
\label{ssc:param_space}
To constrain the amount of chemical processing of \ce{CO} in shielded regions of disks, a grid of chemical and physical conditions typical for disk midplanes inside and outside the \ce{CO} iceline is investigated. The explored parameter range is given in Table~\ref{tab:params}. The disk midplane is assumed to be shielded from stellar and interstellar UV-photons. As such, cosmic-ray induced photons are the only source of UV photons included in the model. The region is also assumed to be shielded from the most intense fluxes of stellar X-rays. The effects of moderate X-ray ionisation rates are similar to the effects of scaled up cosmic-ray ionisation rates \citep{Bruderer2009}. 

There are two steps in this parameter study. First a grid of chemical models for different physical conditions are computed. Temperatures, densities and cosmic-ray ionisation rates typical for cold, shielded regions of protoplanetary disks are used. For these models typical chemical parameters were used. In particular the tunnelling barrier width ($a_\mathrm{tunnel}$) and diffusion-to-binding energy ratio ($f_\mathrm{diff}$), characterizing surface chemistry (see Sec.~\ref{ssc:chem_netw}), were kept constant at 1 {\AA} and 0.3 respectively.

For typical T-Tauri and Herbig disks, the physical conditions probed by our models are shown in Fig.~\ref{fig:Model_over}. The exact specifications of the models are presented in Appendix~\ref{app:Dali_models}. Both models assume a tapered power-law surface density distribution and a Gaussian vertical distribution for the gas. The dust and gas surface densities follow the same radial behaviour with an assumed global gas-to-dust mass ratio of 100. Vertically there is more dust mass near the mid-plane, to simulate dust settling. The temperatures and densities that are included in the chemical models are shown in orange in the bottom panel in Fig.~\ref{fig:Model_over}. The dark orange regions are those that are also completely shielded from VUV radiation. The gas is considered shielded if the intensity of external UV radiation at 100 nm is less than $10^{-4}$ times the intensity at that wavelength from the Draine ISRF \citep{Draine1978,Shen2004}. Our models are however more broadly applicable to other parts of the disks due to dynamical mixing, as described below.

As the T-Tauri disk model is colder (because of the less luminous star) and more compact compared to the Herbig disk model, there is more mass in the region of the disk probed by our models for that disk. The exact extent and location of the region probed by our chemical models strongly depends on the chosen parameters of the disk model. In the Herbig model the temperature never drops below 20 K, as such CO is not frozen out anywhere in that disk model. For selected points in the physical parameter space an additional grid of chemical models is explored (see Sec.~\ref{ssc:Chemical_param}). 

\begin{figure*}
\includegraphics[width = \hsize]{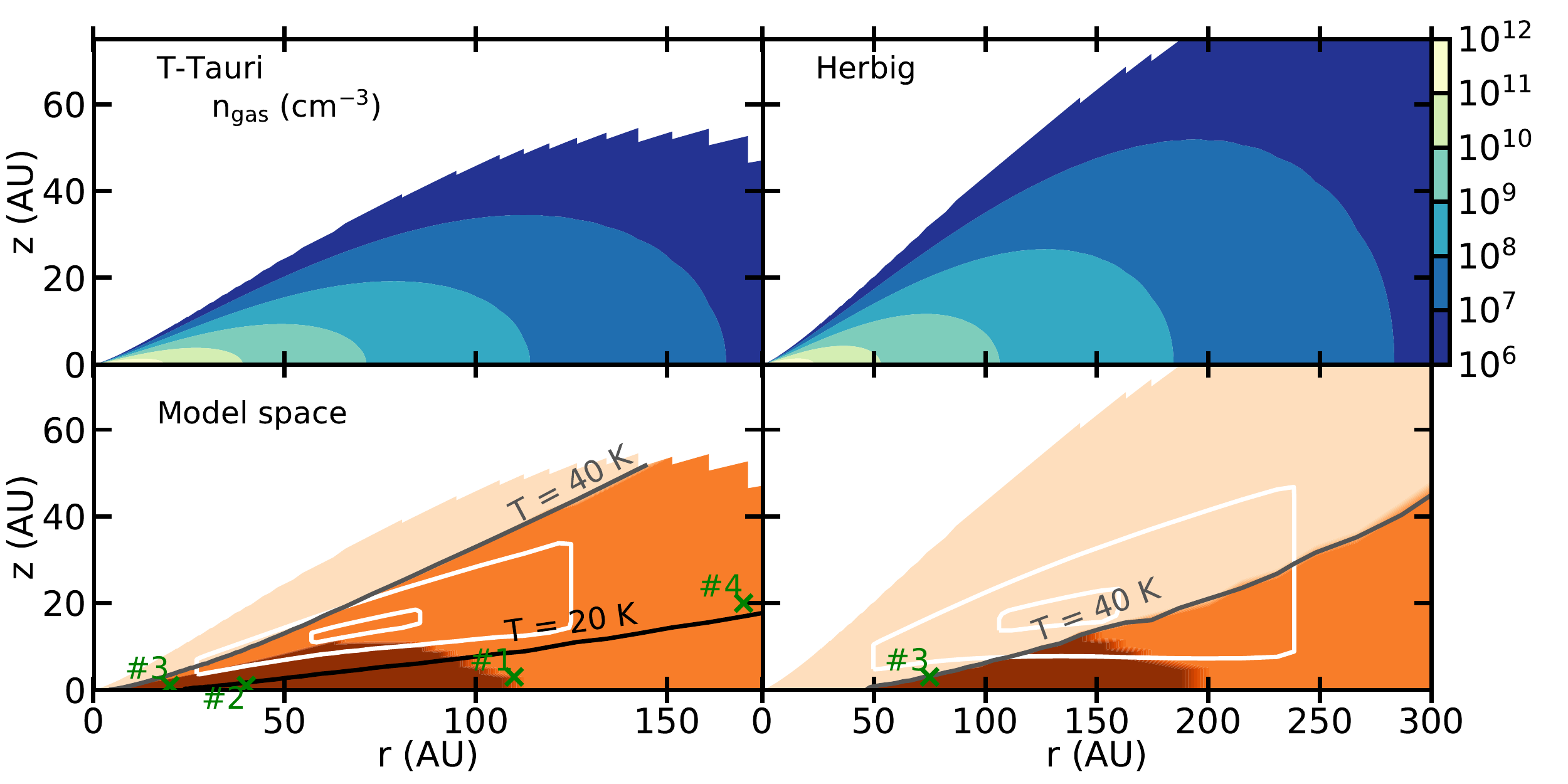}
\caption{\label{fig:Model_over} Number density (\textit{top}) and region of the disk included in the parameter study (\textit{bottom}) for a typical T-Tauri ($L= 0.3$ L$_\odot$) (\textit{left}) and Herbig ($L=20$ L$_\odot$) (\textit{right}) disk model (see App.~\ref{app:Dali_models} for details). The part of the disk that is included in the physical models is shown in dark orange and is bound by the highest temperature included (40 K) and the restriction that the UV is fully shielded. Light orange denotes the region with temperatures and densities probed by our models but with some low level of UV. White contours show the regions contributing to 25\% and 75\% of the emission from the \ce{C^{18}O} 3--2 line. The green crosses numbered \#1, \#2, \#3 and \#4 are approximate locations of the representative models of Sec.~\ref{sssc:rep_mod}.}
\end{figure*}

\subsection{Chemical network}
\label{ssc:chem_netw}
The chemical network used in this work is based on the chemical model from \cite{Walsh2015} as also used in \cite{Eistrup2016,Eistrup2018}. The "\textsc{Rate12}" network from the UMIST Database for Astrochemistry forms the basis of the gas-phase chemistry \citep{McElroy2013}\footnote{Which is available at: \url{http://www.udfa.net}}. \textsc{Rate12} includes gas-phase
two-body reactions, photodissociation and photoionisation, direct cosmic-ray ionisation, and cosmic-ray-induced photodissociation and ionisation. Three-body reactions are not included as they are not expected to be important at the densities used in this work. Photo- and X-ray ionisation and dissociation reactions are included in the network but their contribution is negligible because we assume the disk midplane is well shielded from all external sources of X-ray and UV photons. 

Freeze-out (adsorption) onto dust grains and sublimation (desorption) of molecules is included. Molecules can desorb either thermally or via cosmic-ray induced photodesorption \citep{Tielens1982,Hasegawa1992,Walsh2010,Walsh2012}. Molecular binding energies as compiled for \texttt{Rate12} \citep{McElroy2013} are used, updated with the values recommended in the compliation by \cite{Penteado2017}, except for \ce{NH}, \ce{NH2}, \ce{CH}, \ce{CH2} and \ce{CH3}. For \ce{NH} and \ce{NH2} we calculate new estimates using the formalism proposed by \cite{Garrod2006} and the binding energy for \ce{NH3} \citep[3130 K, ][]{Brown2007}. For the \ce{CH_x} radicals the binding energy is scaled by the number of hydrogen atoms with the \ce{CH4} binding energy of 1090 K as reference \citep{He2014}. A list of all the binding energies used in this work is given in Table~\ref{tab:Bindingenergies}. The binding energy used for \ce{H2}, 430 K, predicts complete freeze-out of \ce{H2} at temperatures up to 15 K at densities of $10^{12}$ cm$^{-3}$. However, at similar densities, \ce{H2} freezes-out completely at much lower temperatures \citep{Cuppen2007}. The binding energy used here is the \ce{H2} to \ce{CO} binding energy, whereas the \ce{H2} to \ce{H2} binding energy is expected to be much lower \citep{Cuppen2007}. As such we modify the binding energy of \ce{H2} such that it is 430 K as long as there is less than one monolayer of \ce{H2} ice on the grain. Above two monolayers of \ce{H2} ice we use the \ce{H2} on \ce{H2} binding energy of 100 K. Between these two regimes, the binding energy of \ce{H2} is linearly dependant on the coverage of \ce{H2} ice. This is a different approach compared to that described in \cite{Hincelin2015} and \cite{Wakelam2016} but it has a similar effect on the \ce{H2} ice abundance. In all cases $E_\mathrm{diff} = f_\mathrm{diff} \times 430 \mathrm{K}$ for the diffusion of \ce{H2}. 

Experimentally determined photodesorption yields are used where available \citep[see, e.g.][] {Oberg2009CO_N2_CO2,Oberg2009H2O,Oberg2009CH3OH}, specifically $2.7 \times 10^{-3}$ \ce{CO} molecules per photon is used from \cite{Oberg2009CO_N2_CO2}. We note that a large range of CO photodesorption yields, between $4\times 10^{-4}$ and $0.25$ \ce{CO} molecules per photon, are available in the literature due to the significant effects of experimental conditions \citep{Oberg2007, Oberg2009CO_N2_CO2, MunozCaro2010, Fayolle2011,Chen2014, MunozCaro2016, Paardekooper2016}. \cite{Fayolle2011} show that temperature and the wavelength of the incident photon strongly influence the photodesorption yield. For all species without experimentally determined photodesorption yields, a value of $10^{-3}$ molecules photon$^{-1}$ is used. 
The sticking efficiency is assumed to be 1 for all species except for the atomic hydrogen that leads to \ce{H2} formation. 

The formation of \ce{H2} is implemented following \cite{Cazaux2004} (see Appendix~\ref{app:H2form} for a summary). This formalism forms \ce{H2} directly out of gas-phase \ce{H}. This fraction of atomic hydrogen is not available for reactions on the grain surface. About 50\% of the atomic hydrogen is used to form \ce{H2} is this way. The remaining atomic hydrogen freezes-out on the grain surface and participates in the grain surface chemistry. Using this formalism ensures that the abundance of atomic \ce{H} does not depend on the adopted grain-surface parameters. The balance between \ce{H2} formation and \ce{H2} destruction by cosmic-rays produces an atomic \ce{H} abundance in the gas that will always be around 1 cm$^{-3}$ independent of the total \ce{H} nuclei abundance. 

For the grain-surface reactions we use the reactions included in the Ohio State University (OSU) network\footnote{http://faculty.virginia.edu/ericherb/research.html} \citep{Garrod2008gas-grain}. The gas-phase network is supplemented with reactions for important chemicals, for example the \ce{CH3O} radical, that are not included in \textsc{Rate12}. The destruction and formation reactions for these species are taken from the OSU network. The grain-surface network also includes additional routes to water formation as studied by \cite{Cuppen2010} and \cite{Lamberts2013}.  The grain-surface reactions are calculated assuming the Langmuir-Hinshelwood mechanism. Only the top two layers of the ice are chemically "active" and we assume that the chemically active layers have the same composition as the bulk ice. No reaction-diffusion competition for grain-surface reactions with a reaction barrier is included \citep{Garrod2011}. The exact equations used to calculate the rates can be found in Appendix~\ref{app:grainsurf}.

The rates for the grain-surface reactions greatly depend on two quantities, the tunnelling barrier ($a_\mathrm{tunnel}$) and the diffusion-to-binding energy ratio ($f_\mathrm{diff}$). $a_\mathrm{tunnel}$ is usually taken to lie between 1 and 1.5 {\AA} \citep{Garrod2011, Walsh2015, Eistrup2016}, and we test the range between 0.5 {\AA} to 2.5 {\AA}. The diffusion-to-binding energy ratio is generally taken to range between 0.3 and 0.5 \citep{Walsh2015, Cuppen2017}, although recent quantum chemical calculations predict values as low as $f_\mathrm{diff} = 0.15$ for \ce{H} on crystalline water ice \citep{Senevirathne2017}. On the other hand, recent experiments suggest fast diffusion rates for CO on \ce{CO2} and \ce{H2O} ices \citep{Lauck2015, Cooke2018}. The range tested here, $f_\mathrm{diff} = 0.1$ to $f_\mathrm{diff} = 0.5$, encompasses this measured range. 

The chemical models are initialised with molecular abundances. The full list of abundances is given in Appendix~\ref{App:init_abu}). Fully atomic initial conditions are not investigated. 

\begin{table*}
\centering
\caption{\label{tab:params} Physical and chemical parameters explored}
\begin{tabular}{l c c c}
\hline \hline
Parameter & Symbol &  Range & Fiducial value\\
\hline
Temperature & $T$ & $10-40$ K & -- \\
Density & $n$ & $10^{6} - 10^{12}$ cm $^{-3}$ & -- \\
Cosmic-ray ionisation rate & $\zeta_{\ce{H2}}$ & $10^{-18} -10^{-16}$ s$^{-1}$ & -- \\
Tunnelling barrier width & $a_\mathrm{tunnel}$ & $0.5-2.5\ \mathrm{\AA}$ & 1 \AA \\
Diffusion-to-binding energy ratio & $f_\mathrm{diff}$ & $0.1-0.5$ & 0.3 \\
\hline
\end{tabular}
\end{table*}

\subsection{CO destruction routes}
\begin{figure*}
\centering
\includegraphics[width = \hsize]{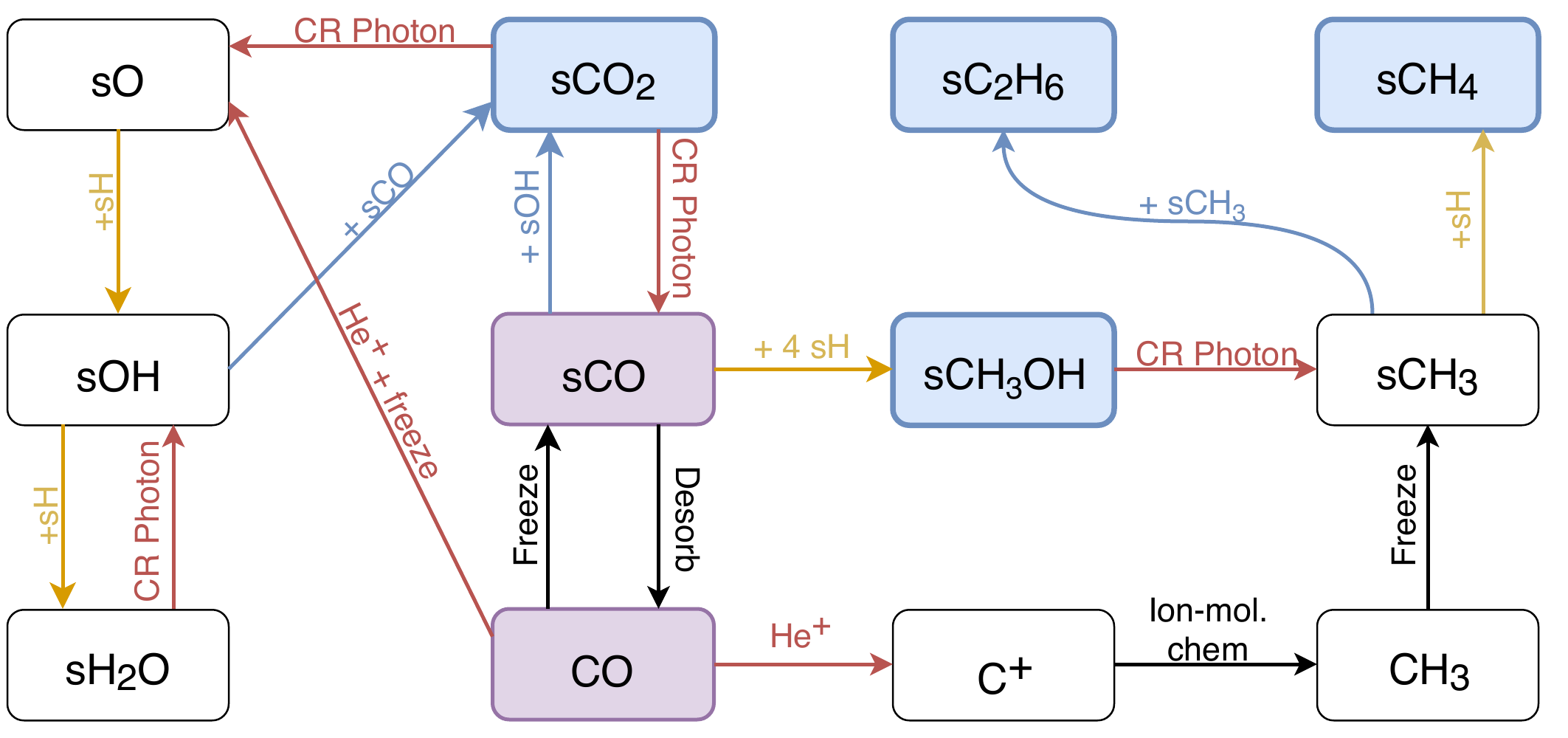}
\caption{\label{fig:CO_destr_schem} Chemical reaction network showing the major \ce{CO} destruction pathways and important competing reactions. Red arrows show reactions that are mediated directly or indirectly by cosmic-ray photons, yellow arrows show hydrogenation reactions and blue arrows show grain-surface reactions. The initial major carbon carrier (\ce{CO}) is shown in grey-purple. Stable products of \ce{CO} processing are denoted with blue boxes. \ce{sX} denotes that species \ce{X} is on the grain surface. }
\end{figure*}

There are three main pathways to destroy CO (see introduction). These are:
\begin{enumerate}
\item \ce{sCO + sH -> sHCO}, leading to \ce{sCH3OH}
\item \ce{sCO + sOH -> sCO2 + sH}
\item \ce{CO + He+ -> C+ + O + He}, leading to \ce{CH4} and \ce{C2H6}
\end{enumerate}
where sX denotes that species X is on the grain-surface. The interactions of these reactions with each other and the major competing reactions are shown in Fig.~\ref{fig:CO_destr_schem}. For each of these reactions a short analysis on the resulting rates is presented to explain the behaviour of the \ce{CO} abundance as shown in Sec.~\ref{sec:Results} using the rate coefficients derived in Appendix~\ref{app:grainsurf}. Table~\ref{tab:chem_lookup} gives an overview of the symbols used, whereas Table \ref{tab:chem_par} shows the sensivity to assumed parameters. 

\begin{table*}
\centering
\caption{\label{tab:chem_lookup} Symbol overview for the rate equations}
\begin{tabular}{c c c}
\hline \hline
Symbol & Parameter & Value [Unit] \\ 
\hline 
$R_{\ce{X + Y}}$ & {\ce{X + Y}} reaction rate & [cm$^{-3}$ s$^{-1}$] \\
$n_\mathrm{X, A}$ & number density of species \ce{X} in phase A & [cm$^{-3}$] \\
$N_\mathrm{sites}$ & number of molecules per layers per grain & $10^6$ \\ 
$n_\mathrm{grain}$ & number density of grains & $2.2\times10^{-12} n_\mathrm{gas}$ [cm$^{-3}$] \\
$C_\mathrm{grain}$ & ice mantle dependent prefactor & $\min\left[1, \left(\frac{N^2_\mathrm{act}N^2_\mathrm{sites}n^2_\mathrm{grain}}{n^2_\mathrm{ice}}\right)\right]/(N_\mathrm{sites}n_\mathrm{grain})$[cm$^{3}$] \\
$\mu$ & reduced mass of the reacting species & [g] \\
$E_\mathrm{bar}$ & height of the reaction barrier & [erg]  \\
$E_\mathrm{bind, X}$ & binding energy of species \ce{X} & [erg] \\
$\nu_\mathrm{X}$ & vibration frequency of species \ce{X} & Eq.~\ref{eq:vibfreq}[s$^{-1}$]\\
\hline
\end{tabular}
\end{table*}

The reaction: 
\begin{equation}
\ce{sCO + sH -> sHCO}
\end{equation}
happens on the surface and through further hydrogenation of \ce{sHCO} leads to the formation of \ce{sCH3OH}. The initial step in this process is the most important and rate limiting step. This reaction has a barrier, $E_\mathrm{bar} = 2500$ K \citep{Woon2002}, which makes tunnelling of \ce{H} very important. The total \ce{CO} destruction rate, assuming that tunnelling dominates for the reaction barrier, and that the thermal rate dominates for \ce{H} hopping, can be given by:
\begin{equation}
\label{eq:CO+H}
\begin{split}
R_\mathrm{CO + H} &= \frac{n_\mathrm{CO, ice} n_\mathrm{H, ice}}{n_\mathrm{CO, total}} C_\mathrm{grain} \exp\left[-\frac{2 a_{\mathrm{tunnel}}}{\hbar}\sqrt{2\mu E_\mathrm{bar}}\right] \times\\
& \left(\nu_\mathrm{H} \exp\left[-\frac{f_{\mathrm{diff}} E_\mathrm{bind, H}}{kT} \right] + \nu_\mathrm{CO} \exp\left[-\frac{f_{\mathrm{diff}} E_\mathrm{bind, CO}}{kT} \right] \right).
 \end{split}
\end{equation}
As noted above, the tunnelling barrier, $a_\mathrm{tunnel}$ is the most important parameter for determining the rate. Changing $a_\mathrm{tunnel} = 0.5$ {\AA} to $a_\mathrm{tunnel} = 1.5$ {\AA} decreases the destruction rate through hydrogenation by eight orders of magnitude. This reaction is also suppressed in regions of high temperature where $n_\mathrm{CO, ice}/n_\mathrm{CO, total}$ is low. 

The amount of \ce{H} in the ice is set by the balance of freeze-out of \ce{H} and the reaction speed of \ce{H} with species in the ice. Desorption of \ce{H} is negligible compared with the reaction of \ce{H} with radicals on the grain in most of the physical parameter space explored. As such there is no strong decrease in the rate near the \ce{H} desorption temperature. This also means that the competition of other iceborn radicals for reactions with \ce{H} strongly influences the rates.

At the lowest temperatures the rates are also slightly suppressed as the hopping rate is slowed. The rate is maximal around the traditional \ce{CO} iceline temperature of around 20 K as $n_\mathrm{CO, ice}/n_\mathrm{CO, total}$ is still high, while thermal hopping is efficient. This is especially so for low values of $f_\mathrm{diff}$, increasing the hopping and thus the reaction rate. This reaction does not strongly depend on density since the absolute flux of \ce{H} arriving on grains is nearly constant as function of total gas density and the rest of the rate only depends on the fraction of \ce{CO} that is frozen out, not the total amount. 

The second reaction is the formation of \ce{CO2} through the grain-surface reaction

\begin{equation}
\label{eq:CO2form}
\ce{sCO + sOH -> sCO2 + sH},
\end{equation}
which has a slight barrier of 400 K \citep{Arasa2013}. It competes with the reaction \ce{sCO + sOH -> sHOCO}. We assume that most of the \ce{HOCO} formed in this way will be converted into \ce{CO2} as seen in the experiments \citep{Watanabe2007,Oba2010, Ioppolo2013} and that is also required to explain \ce{CO2} ice observations. As such we suppress the explicit \ce{HOCO} formation channel in our model. 

The reaction rate for reaction \ref{eq:CO2form} is given by:
\begin{equation}
\label{eq:CO+OH}
\begin{split}
R_\mathrm{CO + OH} &= \frac{n_\mathrm{CO, ice} n_\mathrm{OH, ice}}{n_\mathrm{CO, total}} C_\mathrm{grain} \exp\left[-\frac{E_\mathrm{bar}}{kT}\right]  \times \\
& \left(\nu_\mathrm{OH} \exp\left[-\frac{f_{\mathrm{diff}} E_\mathrm{bind, OH}}{kT} \right] + \nu_\mathrm{CO} \exp\left[-\frac{f_{\mathrm{diff}} E_\mathrm{bind, CO}}{kT} \right] \right).
\end{split}
\end{equation}
As \ce{OH} has a high binding energy of 2980 K \citep{He2014}, 
sublimation of \ce{OH} can be neglected. \ce{CO} sublimation is still important even though the rate is again not dependent on the total \ce{CO} abundance. Due to the strong temperature dependence of the reaction barrier and the \ce{CO} hopping rate, this rate is maximal at temperatures just above the \ce{CO} desorption temperature. Finally the reaction rate depends on the \ce{OH} abundance. \ce{OH} in these circumstances is generally created by the cosmic-ray induced photodissociation of \ce{H2O}, which means that \ce{CO2} formation is fastest when there is a large body of \ce{H2O} ice\footnote{Specifically \ce{H2O} in the upper layers of the ice, but by construction our ice mantles are perfectly mixed}. At late times \ce{H2O} can also become depleted, with \ce{CO2} being the major oxygen reservoir, lowering the supply of \ce{OH} at late times. This lowers the \ce{CO2} production rate, and the destruction of \ce{CO2} can increase the \ce{CO} abundance. 

\ce{CO} has competition with several other radicals for the reaction with \ce{OH}. The most important of these is the competition with \ce{H}. At low densities the $x_{\ce{H}}/x_{\ce{CO}}$ is high, so \ce{OH} will mostly react with \ce{H} to reform \ce{H2O}. Similarly when \ce{H} mobility is increased, by assuming very narrow tunnelling barriers, \ce{sCO2} formation will slow down. At high density $x_{\ce{H}}/x_{\ce{CO}}$ is low, and thus the competition for \ce{OH} is won by \ce{CO}.

\begin{table}
\caption{\label{tab:chem_par} Chemical trends with variations in parameters}
\begin{tabular}{l l c c}
\hline
\hline
Parameter &  & \ce{sCO + sOH} & \ce{sCO + sH} \\ 
\hline 
$\uparrow n $ & $\downarrow x_{\ce{H}}\, $ &  $\uparrow$ & $\downarrow$ \\
$\uparrow a_\mathrm{tunnel}$ & $\downarrow$ $P_\mathrm{reac}(\ce{sCO}, \ce{sH})$  & -- & $\downarrow$ \\
$\uparrow a_\mathrm{tunnel}$ & $\downarrow$ $P_\mathrm{reac}(\ce{sOH}, \ce{sH2})$  & $\uparrow$ & -- \\
$\uparrow f_\mathrm{diff}$ & $\downarrow$ CO mobility  & $\downarrow$ & -- \\
\hline
\end{tabular}
\end{table}

The last reaction is the only gas-phase route
\begin{equation}
\ce{CO + He+ -> C+ + O + He}.
\end{equation}
This reaction is limited by the ionisation rate of \ce{He} and the subsequent competition for collisions of \ce{He+} with abundant gas-phase species. As such the \ce{CO} destruction rate can be expressed as:
\begin{equation}
\begin{split}
R_\mathrm{CO + He^+} = 0.65 \zeta_{{H_2}}\frac{x_{\mathrm{{He}}}}{x_{\mathrm{CO}}} 
\frac{k _\mathrm{ion, {CO}}x_{\mathrm{CO}}}{
\sum_\mathrm{X} k_\mathrm{ion, \mathrm{X}}x_\mathrm{X}}
\end{split}
\end{equation}
where $k_\mathrm{ion, \mathrm{X}}$ are the ion-neutral reaction rate coefficients for collisions between \ce{He+} and the molecule and $x_\mathrm{X}$ is the abundance of species $\mathrm{X}$. The abundances and rate coefficients for important alternative reaction partners of \ce{He+} between 20 and 40 K are tabulated in Table~\ref{tab:Hereac}, where we have summed the rate coefficients of reactions with multiple outcomes. At high abundances of \ce{CO} and/or \ce{N2} the rate scales as:
\begin{equation}
R_{\ce{CO + He^+}} \propto \frac{1}{x_{\ce{CO}} + x_{\ce{N2}}},
\end{equation}
which increases with lower abundances. If the sum of the gaseous abundances of \ce{CO} and \ce{N2} is $ << 3 \times 10^{-7}$ the rate becomes
\begin{equation} 
R_{\ce{CO + He^+}} = 0.65 \zeta_{\ce{H2}}x_{\ce{He}} \frac{k_\mathrm{ion, \ce{CO}}}{k_\mathrm{ion, \ce{H2}}x_{\ce{H2}} + k_\mathrm{ion, grains}x_\mathrm{grains}}
\end{equation}
which is independent of the \ce{CO} abundance.

\begin{table}
\caption{\label{tab:Hereac} Rate coefficients for collisions with \ce{He+}}
\begin{tabular}{l c c}
\hline
\hline
Reaction partner & Rate coeff. (cm$^{-3}$ s$^{-1}$) & Gas abundance\\
\hline
\ce{H2} & $1.14 \times 10^{-14}$ & 0.5 \\
\ce{N2} & $1.6 \times 10^{-9} $& $< 2\times 10^{-5}$\\
\ce{CO} & $1.6 \times 10^{-9} $& $< 10^{-4}$\\
grains  & $2.06 \times 10^{-4} $& $2.2\times 10^{-12}$\\
\hline
\end{tabular}
\end{table}

There are some assumptions in the model that will influence the rates of the chemical pathways discussed here. We do not expect the chemistry to be critically dependent on these assumptions but they might influence the chemical timescales and the relative importance of different chemical pathways. A few important assumptions and their effects on the chemistry are discussed in Appendix.~\ref{app:modassump}.

\section{Results}
\label{sec:Results}
We have performed a parameter space study of the chemistry of \ce{CO} under shielded conditions in protoplanetary disks. In this section we first present the results for the physical parameters studied, namely chemical evolution time, density, temperature and cosmic-ray ionisation rate. Fig.~\ref{fig:CR_CO} focuses on the effects of time and cosmic-ray ionisation rate, while Fig.~\ref{fig:dens_temp_chem} focuses on the effects of temperature and density. Together these figures show that the evolution of the CO abundance depends strongly on the physical conditions assumed in the chemical model, especially the temperature, in addition to the cosmic-ray ionisation rate identified earlier. Finally, the effects of the assumed chemical parameters on four positions in physical parameter space are studied.

\subsection{Physical parameter space}
\label{ssc:phys_param}
\subsubsection{Importance of the cosmic-ray ionisation rate}
\label{sssc:COdestr}
\begin{figure*}

\includegraphics[width=\hsize]{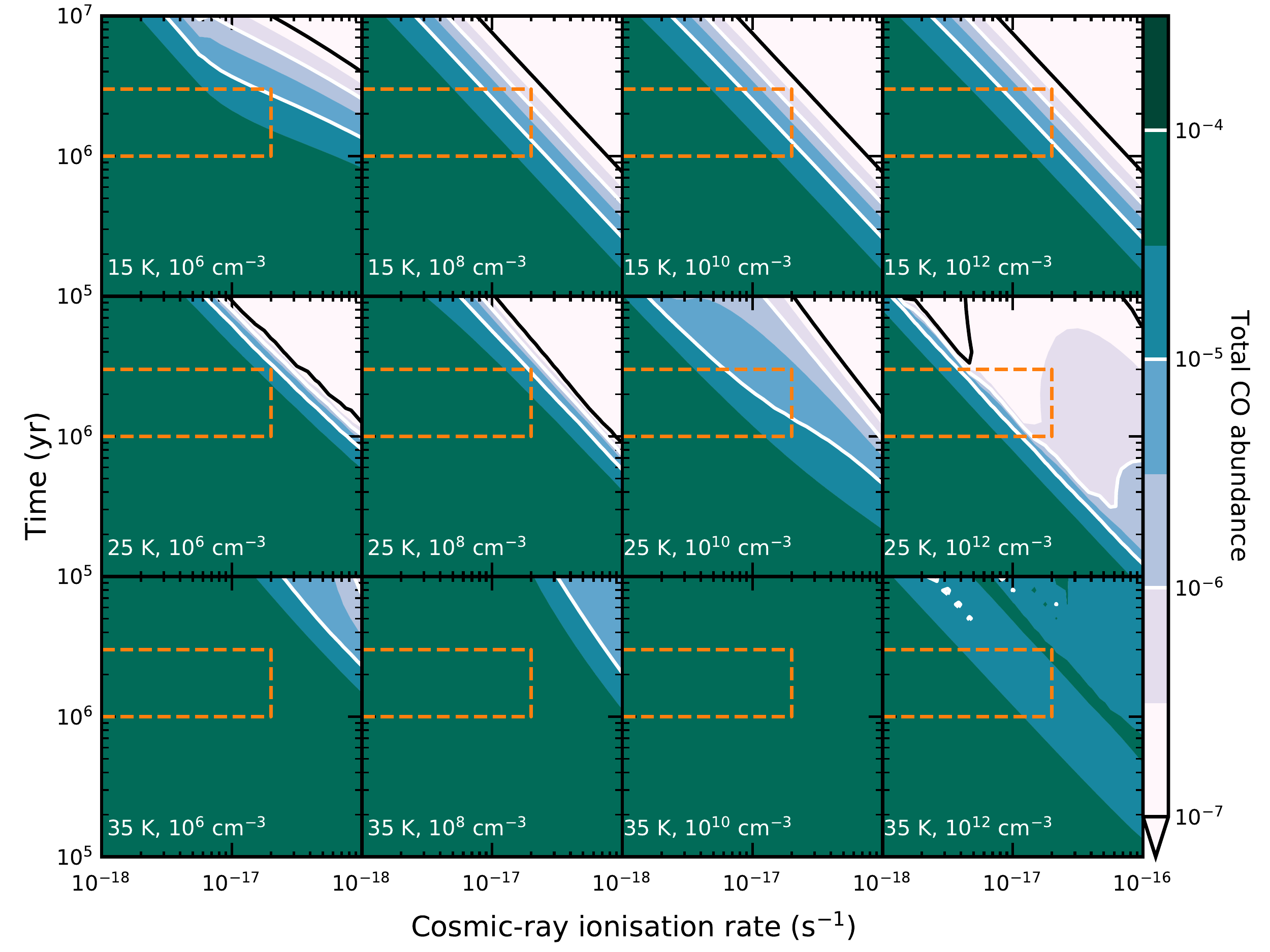}

\caption{\label{fig:CR_CO} Total CO abundance (gas and ice) as function of time and cosmic-ray ionisation rate for the fiducial chemical model. Each sub-figure has a different combination of temperature and density as denoted in the bottom left. Time and cosmic-ray ionisation rate are degenerate in most of the parameter space. The orange box denotes the combinations of $\zeta_{\ce{H2}}$ and time most appropriate for protoplanetary disks. }
\end{figure*}

Consistent with previous studies, the cosmic-ray ionisation rate is found to be the driving force behind most of the changes in the CO abundance. A higher cosmic-ray ionisation rate allows the chemistry to evolve faster, but in a similar way. As such the cosmic-ray ionisation rate and chemical evolution time are mostly degenerate. Fig.~\ref{fig:CR_CO} presents an overview of the dependence of the total CO abundance (gas plus ice) on evolution time and $\zeta_{\ce{H2}}$. \ce{CO} can be efficiently destroyed in 1--3 Myr for $\zeta_{\ce{H2}} > 5 \times 10^{-18}$ s$^{-1}$ and temperatures lower than 25 K. 

For models at 15 K and low densities of $10^6$ cm$^{-3}$, the \ce{CO} abundance behaviour does not show the degeneracy between $\zeta_{\ce{H2}}$ and time. This is caused by the formation of \ce{NO} in the ice. The \ce{NO} abundance depends non-linearly on the cosmic-ray ionisation rate. A high abundance of \ce{NO} in the ice lowers the abundance of available atomic \ce{H} on the ice as it efficiently catalyses the formation of \ce{H2} on the ice. This effect has also been seen in \cite{Penteado2017} using the same network but under different conditions. A similar catalytic effect for the formation of \ce{H2} was first noted in the work by \cite{Tielens1982}.

At high density and temperature (35 K, $10^{10}$ cm$^{-3}$), a sequence of \ce{CO} destruction and then reformation is visible in the \ce{CO} abundance. The first cycle of this was also seen and discussed in \cite{Eistrup2016} and can be attributed to the lower formation rate of \ce{OH} due to a decrease in the \ce{H2O} abundance on Myr timescales. Some models, especially those with a cosmic-ray ionisation rate of $10^{-16}$ s$^{-1}$ can have five or more of these \ce{CO}-\ce{CO2} abundance inversions, while the \ce{H2O} abundance continues to decrease. For the rest of the results in this section a cosmic-ray ionisation rate of $10^{-17}$ s$^{-1}$ is taken, thought to be typical for dense molecular clouds \citep[e.g.,][]{Dalgarno2006}.

\subsubsection{Importance of temperature}

\begin{figure*}
\includegraphics[width = \hsize]{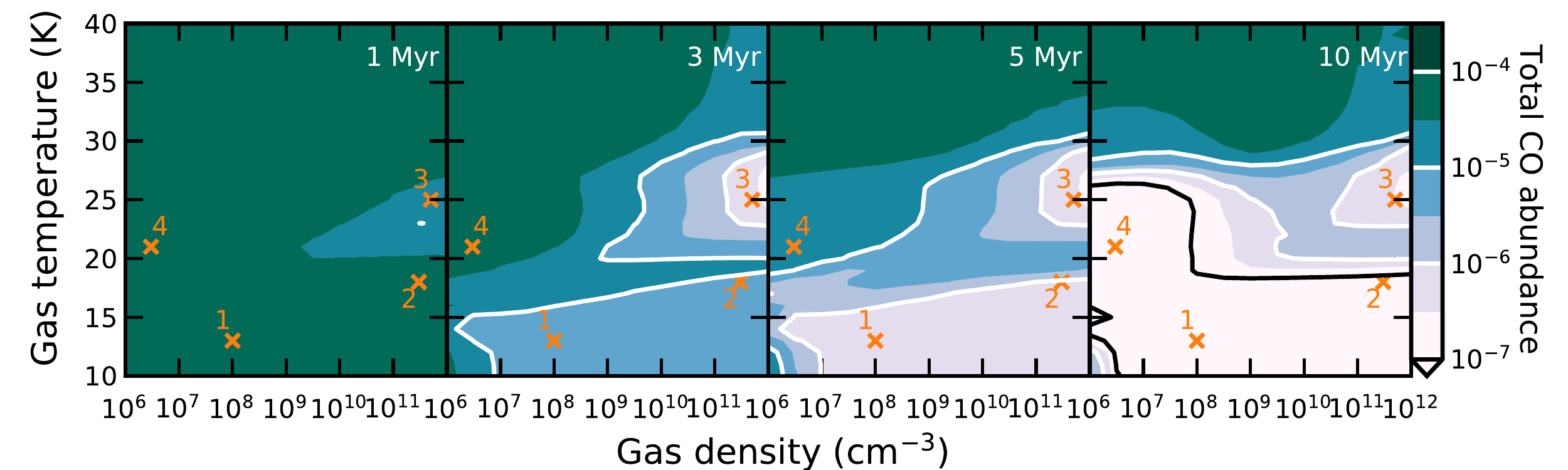}
\caption{\label{fig:dens_temp_chem} Time evolution of \ce{CO} abundance as function of gas temperature and density. Chemical evolution time is denoted in the upper right corner of each panel and $\zeta = 10^{-17}$ s$^{-1}$ for all of these models. The orange numbers show the locations of the physical conditions taken in Sec.~\ref{ssc:Chemical_param}. }
\end{figure*}

Fig.~\ref{fig:dens_temp_chem} presents the total \ce{CO} abundance over the entire density-temperature grid at four time steps during the evolution of the chemistry. These figures demonstrate clearly that \ce{CO} is efficiently destroyed only at low temperatures, $<$30 K. At early times CO is most effectively destroyed at high densities and temperatures between 20 and 30 K. In this range, the grain-surface formation of \ce{CO2} from the reaction between CO and \ce{OH} is efficient. At these temperatures \ce{CO} is primarily in the gas-phase but a small fraction resides on the grain surface where it is highly mobile. \ce{OH} is created during the destruction of \ce{H2O} ice. This reaction is most efficient under high-density conditions because atomic H competes with \ce{CO} for reaction with the \ce{OH} radical on the grain. At low densities, the relative abundance of \ce{H} is higher in the models, thus greatly suppressing the formation rate of \ce{CO2} from \ce{CO + OH} on the grain.  

At 3 Myr \ce{CO} destruction below 20 K becomes visible. The total \ce{CO} abundance in this temperature range is only weakly dependent on the density: CO destruction is efficient below 16 K at the lowest densities, while it is efficient up to 19 K at the highest densities. This range is strongly correlated with the fraction of CO residing in the ice.

At late times, $>5$ Myr, there is a strong additional CO destruction at densities $<10^{9}$ cm$^{-3}$ and at temperatures between 20 and 25 K. At this point the \ce{C+} formed from the \ce{CO + He+} reaction can efficiently be converted into \ce{CH4}, which freezes out on the grains. 

\subsubsection{Representative models}
\label{sssc:rep_mod}
\begin{figure*}
\includegraphics[width=\hsize]{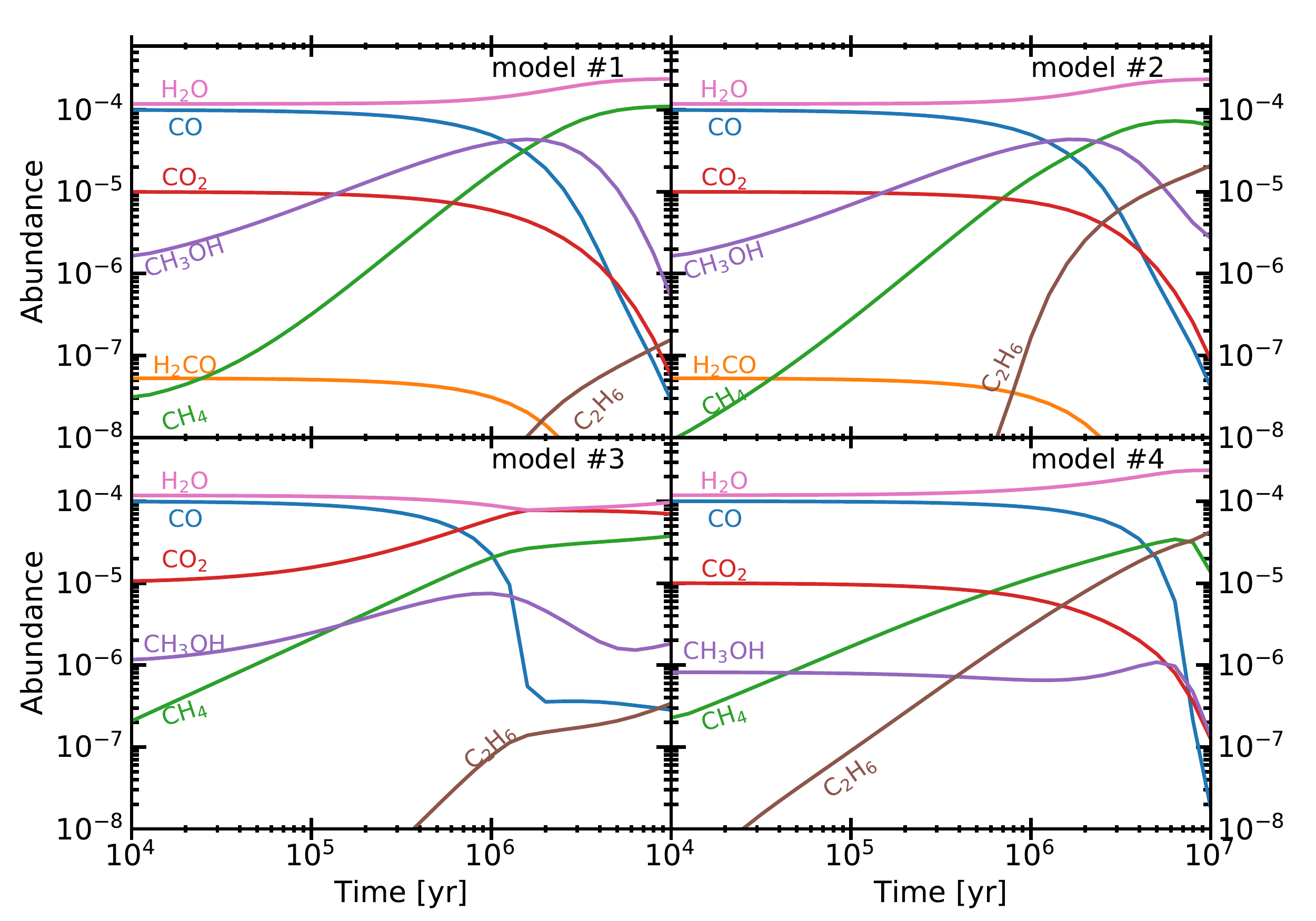}
\caption{\label{fig:abundance_trace} Abundance traces of \ce{CO} and its destruction products for the four points denoted in Fig.~\ref{fig:dens_temp_chem} with conditions as tabulated in Table~\ref{tab:chem_param_test_points}. Plotted abundances are the sum of the gas and ice abundance for each species.}
\end{figure*}

Four points in the physical parameter space have been chosen for further examination. They are chosen such that they span the range of parameters that can lead to low total \ce{CO} abundances within 10 Myr and such that they sample different \ce{CO} destruction routes. These points are given in Table~\ref{tab:chem_param_test_points} and are marked in Fig.~\ref{fig:dens_temp_chem}. Figure~\ref{fig:abundance_trace} presents the total abundance (gas and ice) of \ce{CO} and its stable reaction products as a function of time for the four physical conditions that have been chosen. 

Models \#1 (13 K, $10^8$ cm$^{-3}$) and \#2 (18 K, $3 \times 10^{11}$ cm$^{-3}$) show very similar behaviours, even though they have very different densities. This is due to the combination of the active destruction pathway, \ce{CO} hydrogenation, and the \ce{H2} formation prescription, which forces a constant atomic hydrogen concentration, leading to a constant \ce{CO} destruction rate as a function of density. In both models, 90 \% of the \ce{CO} has been converted into \ce{CH3OH} in 2 Myr. Before this time the \ce{H2CO} abundance is constant, balanced between the formation due to \ce{CO} hydrogenation, and destruction due to hydrogenation. As the \ce{CO} abundance drops, and thus the formation rate of \ce{H2CO} falls, so does its abundance. After slightly more than 2 Myr, methanol has reached a peak abundance close to $10^{-4}$. This marks the end of hydrogenation driven chemical evolution as most molecules on the ice cannot be hydrogenated further. After this, cosmic-ray induced dissociation dominates the abundance evolution, slowly destroying \ce{CH3OH}, forming \ce{CH4} and \ce{H2O}, and destroying \ce{CO2} forming \ce{CO} and \ce{O}, both of which quickly hydrogenate to \ce{CH3OH} and \ce{H2O}. A small amount of \ce{CH4} is further converted into \ce{C2H6}, which happens more efficiently at higher densities, due to the lower availability of atomic \ce{H}.

The abundance traces for model \#3 (25 K, $5 \times 10^{11}$ cm$^{-3}$) show two different destruction pathways for \ce{CO} at a temperature where most of the \ce{CO} is in the gas-phase. The rise in \ce{CO2} abundance indicates that a significant portion of the \ce{CO} reacts with \ce{OH} on the grain-surface to form \ce{CO2}. At 2 Myr nearly 99\% of the initial \ce{CO} has been destroyed. Most of the \ce{CO} has been incorporated into \ce{CO2} with a significant amount of carbon locked into \ce{CH4}. \ce{CH4} is formed from a sequence of reactions that begin with the destruction of \ce{CO} by \ce{He+}. At these late times, \ce{C2H6} acts as a carbon sink, slowly locking up carbon that is created in the form of the \ce{CH3} radical from the dissociation of \ce{CH4}.

The abundance traces for model \#4 (21 K, $3 \times 10^{6}$ cm$^{-3}$) are an outlier in this comparison. Most of the \ce{CO} is in the gas-phase at this temperature and density. It takes at least 2 Myr to destroy 50\% of the \ce{CO} and 5 Myr to destroy 90\% of the \ce{CO}. For the conditions shown here, most of the \ce{CO} is destroyed in the gas-phase by dissociative electron transfer with \ce{He+}. This leads to the formation of hydrocarbons in the gas-phase, primarily \ce{CH3} and \ce{C2H2}, which freeze-out and are hydrogenated on the grain to form \ce{CH4} and \ce{C2H6} \citep[as also seen by][]{Aikawaflow1999}.

\begin{table}
\centering
\caption{\label{tab:chem_param_test_points} Physical parameters for the chemical parameters test.}
\begin{tabular}{c c c c}
\hline
\hline
Model \# & $n_{\mathrm{gas}}$ (cm$^{-3}$) & $T_{\mathrm{gas}}$ (K) & $\zeta_{\ce{H2}}$ (s$^{-1}$)\\ 
\hline
1  & $ 1 \times 10^8$ & 13 &  $10^{-17}$\\
2  & $ 3 \times 10^{11}$ & 18 &  $10^{-17}$\\
3  & $ 5 \times 10^{11}$ &  25 & $10^{-17}$\\
4  & $ 3\times 10^6$ & 21 &  $10^{-17}$\\
\hline
\end{tabular}
\end{table}
\subsection{Chemical parameter space}
\label{ssc:Chemical_param}
For the four different cases listed in  Table~\ref{tab:chem_param_test_points}, a set of models with varying $a_\mathrm{tunnel}$ and $f_\mathrm{diff}$ have been computed. $a_\mathrm{tunnel}$ primarily changes the reaction probability for grain-surface reactions involving atomic or molecular hydrogen that have a barrier, such as \ce{sCO + sH} and \ce{sOH + sH2}. The value of $f_\mathrm{diff}$ changes the speed at which species can move over the grain-surface. Models \#1 and \#2 are both in the region of parameter space where \ce{CO} is frozen out and thus sample pure grain-surface chemistry at low temperature and density, and at a slightly higher temperature and high density respectively. Model \#3 is near the local minimum in gaseous CO abundance seen in Fig \ref{fig:dens_temp_chem}. Model \#4 is located in a region of parameter space where most changes are still ongoing at later times. Together these four cases should sample the different dominant \ce{CO} destruction pathways.

\begin{figure*}
\includegraphics[width = \hsize]{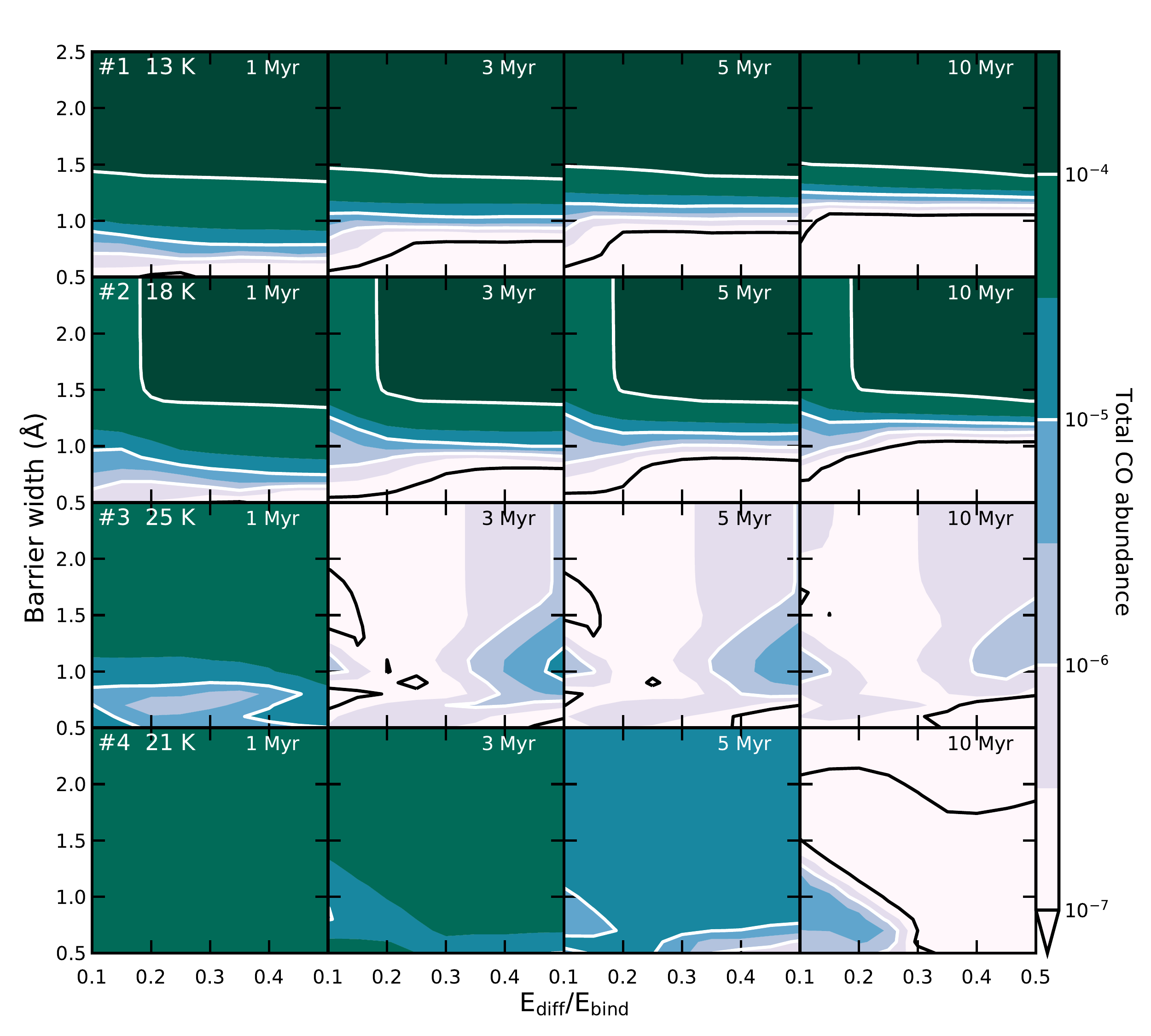}
\caption{\label{fig:All_chemstudy} Time evolution of total \ce{CO} abundance (ice and gas) as function of the assumed tunnelling barrier width ($a_\mathrm{tunnel}$) and diffusion-to-binding energy ratio ($f_\mathrm{diff}$). All these models have $\zeta_{\ce{H2}} = 10^{-17}$ s$^{-1}$. The first row of models uses $T_\mathrm{gas} = 13$ K, $n_\mathrm{gas} = 10^{8}$ cm$^{-3}$. The second row of models uses $T_\mathrm{gas} = 18$ K and $n_\mathrm{gas} = 3 \times 10^{11}$ cm$^{-3}$. The third row of models uses $T_\mathrm{gas} = 25$ K and $n_\mathrm{gas} = 5 \times 10^{11}$ cm$^{-3}$. The fourth row of models uses $T_\mathrm{gas} = 21$ K and $n_\mathrm{gas} = 3 \times 10^{6}$ cm$^{-3}$. Significant reduction of the total CO abundance in less than 3 Myr is only possible if \ce{sCO + sH} is efficient, which is at low barrier widths or if \ce{sOH} preferably reacts with \ce{sCO}, which is enhanced at low values of $f_\mathrm{diff}$.The fiducial values are $a_\mathrm{tunnel}=1 \AA$ and $f_\mathrm{diff}=0.3$). }
\end{figure*}

The first row of Fig.~\ref{fig:All_chemstudy} shows the total \ce{CO} abundance as function of chemical parameters for different times for point \#1 in our physical parameter space with \ce{CO} mostly in the ice. There is a strong dependence on the tunnelling barrier width ($a_\mathrm{tunnel}$). This is because the \ce{CO} destruction in this temperature regime is dominated by the formation of \ce{sHCO}. The \ce{sCO + sH} reaction has a barrier which strongly limits this reaction, \ce{H} tunnelling through this barrier thus increases the rate of \ce{CO} destruction. Our fiducial value for the barrier width, 1 \AA , is in the transition between slow hydrogenation at higher barrier widths and runaway hydrogenation at lower barrier widths. As such a 10\% change to the barrier width around 1 \AA{} changes the abundance of \ce{CO} by a factor of $\sim 3$.

There are only very weak dependencies on the diffusion-to-binding energy ratio ($f_\mathrm{diff} = E_\mathrm{diff}/E_\mathrm{bind}$) for these very low temperatures. This points at a CO destruction process that is entirely restricted by the tunnelling efficiency of \ce{H}. If the \ce{sCO + sH} reaction is quenched by a large barrier, \ce{CO} destruction is so slow that, due to the destruction of \ce{CO2} by cosmic-ray induced photons, the \ce{CO} abundance is actually increased from the initial value. This happens at barrier widths larger than 1.5 {\AA}. At the lowest $f_\mathrm{diff}$ \ce{CO} is turned into \ce{CO2} through \ce{sCO + sOH} at early times. At later times, the \ce{CO2} is destroyed, again by cosmic-ray induced photons and more \ce{CO} is formed, leading to a slower \ce{CO} abundance decrease at late times. 

The second row of Fig.~\ref{fig:All_chemstudy} shows the total \ce{CO} abundance as a function of chemical parameters for different times for point \#2. Since \ce{CO} is frozen out for both case \#1 ($T=13$ K) and \#2 ($T=18$ K), there are strong similarities between the first and second row of models in Fig.~\ref{fig:All_chemstudy}. The only significant difference can be seen at $a_\mathrm{tunnel} > 1.0 \mathrm{\AA}$ and $f_\mathrm{diff} < 0.2$. With these chemical parameters and these high densities ($3\times 10^{11}$ cm$^{-3}$), \ce{sCO + sOH} can be an effective destruction pathway.

The third row of Fig.~\ref{fig:All_chemstudy} shows the total \ce{CO} abundance for point \#3 in our physical parameter space at $T = 25$ K and $n = 5\times 10^{11}$ cm$^{-3}$. The large and irregular variation in \ce{CO} abundance in this figure points at a number of competing processes. The two main processes destroying \ce{CO} at this temperature and density are again \ce{sCO + sH} and \ce{sCO + sOH}. These grain-surface reactions dominate the CO destruction even though \ce{CO} is primarily in the gas-phase. The fraction of time that \ce{CO} spends on the grain is long enough to allow the aforementioned reactions to be efficient. At the smallest barrier widths (< 0.7 {\AA}) the conversion of \ce{CO} into \ce{CH3OH} through hydrogenation dominates the \ce{CO} abundance evolution, but this pathway quickly gets inefficient if the tunnelling barrier is made wider. In the rest of the parameter space the formation of \ce{CO2} dominates the abundance evolution.

Although CO is destroyed significantly in this model from 2 Myr onward, there is a clear region in parameter space where the \ce{CO} destruction is slower. This region at high $f_\mathrm{diff}$ and intermediate $a_\mathrm{tunnel}$ has up to two orders of magnitude higher \ce{CO} abundance compared with the rest of parameter space. The high $f_\mathrm{diff}$ significantly slows down all reactions that are unaffected by tunnelling. In this region of parameter space, CO mobility is significantly lower than \ce{H} mobility due to the latter being able to tunnel. This suppresses the \ce{sCO + sOH} route. On the other hand the barrier is still too wide to allow efficient hydrogenation of \ce{CO}. With both main destruction routes suppressed in this region, it takes longer to reach a significant amount of \ce{CO} destruction. Further increasing $a_\mathrm{tunnel}$ from this point slows down the tunnelling of \ce{H} and thus the formation rate of \ce{H2O}. This leads into a larger \ce{OH} abundance on the ice, and thus a larger \ce{CO2} formation rate.

The fourth row of Fig.~\ref{fig:All_chemstudy} shows the total \ce{CO} abundance at different times for point \#4 at a low density of $3\times 10^6$ cm$^{-3}$ and $T=21$ K. At early times there is no strong destruction of \ce{CO}, after 3 Myr there is only a small region where the abundance has dropped by a factor of three. \ce{CH3OH} and \ce{CO2} are formed in this region of parameter space. At 5 Myr, there is a region at low barrier width, around 0.6 {\AA} and $f_\mathrm{diff} > 0.3$, where the hydrogenation of \ce{CO} has led to a large decrease in the \ce{CO} abundance. At 7 Myr this process has caused a decrease of \ce{CO} abundance of at least two orders of magnitude in this region. At the same time the \ce{CO} abundance over almost all of the parameter space has dropped by an order of magnitude. This is the effect of the gas-phase route starting with dissociative electron transfer of \ce{CO} to \ce{He+}: \ce{CO + He+ -> C+ + O + He}. In the last 3 Myr of chemical evolution, this reaction pathway removes 99.9 \% of the \ce{CO} in a large part of the parameter space. Only regions that had a significant build up of \ce{CO2} at early times, or where \ce{N2} can be efficiently reformed, show \ce{CO} abundances $> 10^{-6}$, because \ce{CO} is reformed from \ce{CO2} and because \ce{N2} competes with \ce{CO} for reactions with \ce{He+} respectively. 

In summary, our results do not strongly depend on the value of $f_\mathrm{diff}$, however the speed of hydrogenation critically depends on the value for $a_\mathrm{tunnel}$, especially around $a_\mathrm{tunnel}=1 \AA$.
Several independent
laboratory experiments show that the CO hydrogenation proceeds fast,
even at temperatures as low as 10--12 K \citep{Hiraoka2002,Watanabe2002,Fuchs2009} so a high barrier is unlikely. Reaction probabilities from the Harmonic Quantum Transition State calculations by \cite{Andersson2011} are consistent with $a_\mathrm{tunnel} \approx 0.9 \AA$ in our calculations. Given the importance of the exact value of the tunnelling barrier on the chemical evolution, more work is needed in understanding the tunnelling of hydrogen during hydrogenation processes.

\section{Discussion}
Our results show that it is possible to chemically process \ce{CO}
under conditions ($T < 30$ K, $n = 10^6-10^{12}$ cm$^{-3}$) that are
representative of a large mass fraction of a protoplanetary disk on a
few Myr timescale. As such, chemical processing of \ce{CO},
specifically the formation of \ce{CH3OH} and \ce{CO2} and on longer
timescales hydrocarbons, has a significant effect on the observed
\ce{CO} abundance.

Our results agree with \cite{Reboussin2015}, \cite{Eistrup2016} and
\cite{Schwarz2018} that a significant cosmic-ray ionisation rate, $>
5\times 10^{-18}$ s$^{-1}$ in our study, is needed to convert \ce{CO}.

In contrast with \cite{Furuya2014} and \cite{Yu2016}, we do not find that destruction of \ce{CO} through reactions with \ce{He+} is a main pathway for a cosmic-ray ionisation rate of $10^{-17}$ s$^{-1}$. This is mostly because this destruction timescale is $>5$ Myr, for these levels of ionisation. However, \cite{Furuya2014} assume a higher cosmic-ray ionisation rate ($5\times 10^{-17} $s$^{-1}$) whereas \cite{Yu2016} have X-rays that add to the total ionisation rate of the gas. This lowers their \ce{CO} destruction time-scale significantly (Fig.~\ref{fig:CR_CO}).

\subsection{When, where and how is CO destroyed within 3 Myr}
\begin{figure*}
\centering
\includegraphics[width = \hsize]{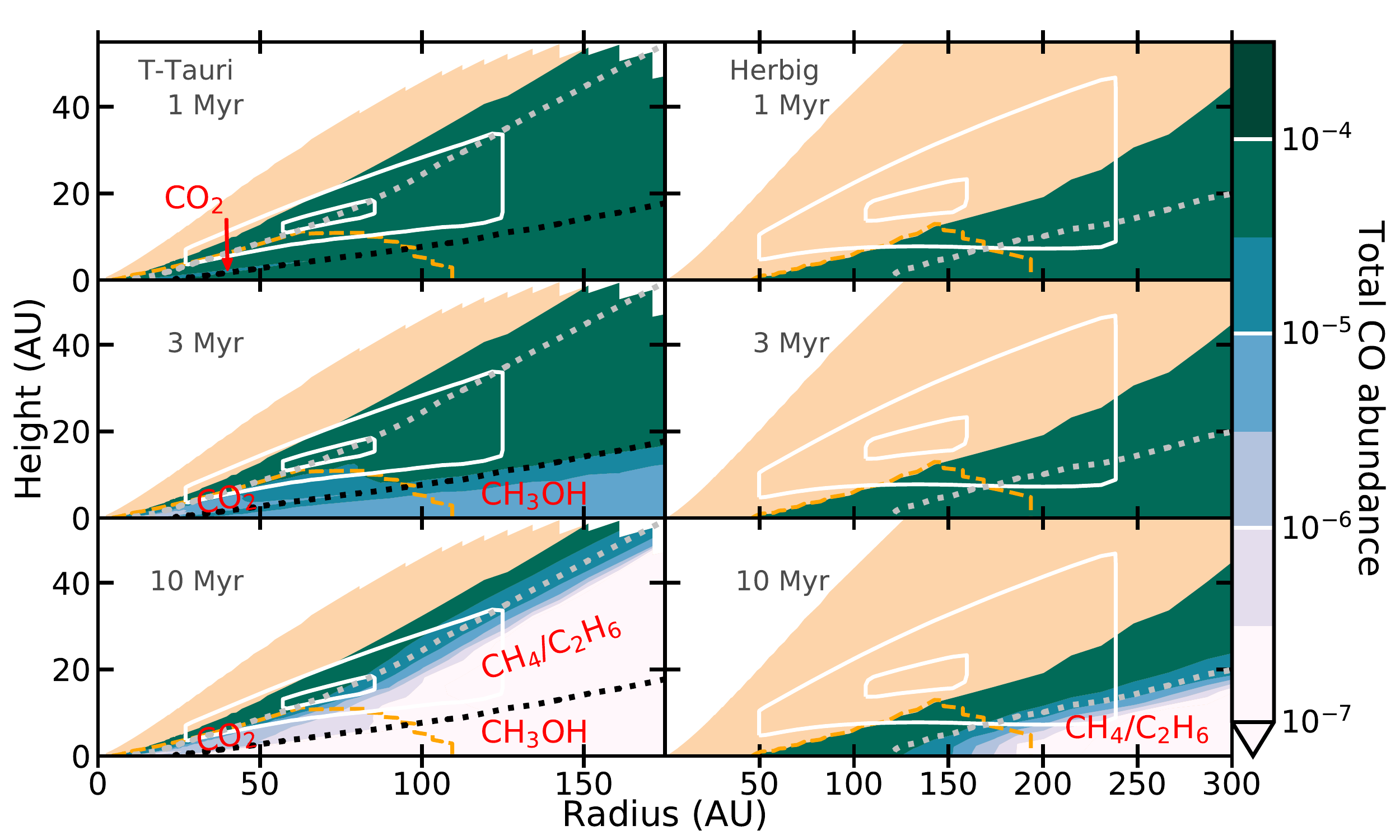}
\caption{\label{fig:COmapped} Total CO abundance (gas and ice) from the chemical models, at 1, 3 and 10 Myr, mapped onto the temperature and density structure of a T-Tauri and Herbig disk model (see Fig.~\ref{fig:DALI_full} and Appendix~\ref{app:Dali_models}). A cosmic-ray ionisation rate of $10^{-17}$ s$^{-1}$ is assumed throughout the disk. The orange hatched area shows the disk region that has a temperature above 40 K and is thus not included in our chemical models. See Fig.~\ref{fig:DALI_full} for the CO abundance in that region using a simple CO chemistry. The white contours shows the area within which 25\% and 75\% of the \ce{C^{18}O} flux is emitted. The black and grey dotted lines show the 20 K and 30 K contours respectively. The Herbig disk is always warmer than 20 K. The molecules in red denote the dominant carbon carrier in the regions \ce{CO} is depleted. The orange dashed line encompasses the region that is completely shielded from UV radiation (see Sec.~\ref{ssc:param_space}). }
\end{figure*}

There are two main reaction mechanisms that can destroy \ce{CO} in 3
Myr under cold disk midplane conditions, assuming a cosmic-ray rate of
$10^{-17}$ s$^{-1}$: hydrogenation of \ce{CO} to \ce{CH3OH} and the
formation of \ce{CO2}.  Hydrogenation of CO forming \ce{CH3OH} reduces
\ce{CO} by an order of magnitude within 2 Myr in the cold regions of
the disk where \ce{CO} is completely frozen out
(Fig.~\ref{fig:abundance_trace}, top two panels; Fig.~\ref{fig:COmapped}, left column, top two panels). The first step of
this pathway \ce{sCO + sH} has a barrier and thus the time-scale of this
reaction is strongly dependent on the assumed hydrogen tunnelling
efficiency ($a_\mathrm{tunnel}$), with shorter timescales at lower
assumed $a_\mathrm{tunnel}$ (Fig.~\ref{fig:All_chemstudy}).

Contrary to \cite{Schwarz2018} we find that \ce{CH3OH} formation is
only dominant when \ce{CO} is frozen-out, $<$20 K. This is probably
due to the more complete grain-surface chemistry in our work, allowing
other species to react with atomic \ce{H} and removing it from the
grain-surface before \ce{CO} can react with it at higher
temperatures. This leads us to conclude that \ce{CO2} formation is
always the dominant \ce{CO} destruction mechanism in the warm
molecular layer in the first 5 Myr instead of \ce{CO} hydrogenation, showing that a
larger grain-surface network needs to be considered.

The second pathway, \ce{CO + OH -> CO2} on the ice, is efficient at
slightly higher temperatures, between the \ce{CO} iceline temperature
and up to 35 K at the highest densities
(Fig.~\ref{fig:dens_temp_chem}). This pathway leads to $\sim$2 orders decrease in the \ce{CO} abundance
 within 3 Myr. The rate of conversion of \ce{CO} into
\ce{CO2} depends strongly on the assumed chemical parameters in the
first Myr. By 3 Myr most of these initial differences have been washed
out. Only a very slow assumed diffusion rate ($f_\mathrm{bind} >
0.35$) leads to a \ce{CO} abundance above $10^{-6}$ after 3 Myr (Fig.~\ref{fig:All_chemstudy}).

Production of hydrocarbons follows the formation of \ce{CH3OH} in the first Myr. When the \ce{CO} abundance has been lowered by 1 to 2 orders of magnitude, production of \ce{CH3OH} stops, while the formation of \ce{CH4} continues. At $\sim 3$ Myr \ce{CH4} becomes more abundant than \ce{CH3OH}. At even longer timescales, \ce{CH4} gets turned into \ce{C2H6}. Formation of hydrocarbons starting from \ce{CO + He^+} in the gas phase is only effective at densities $< 10^9$ cm$^{-3}$ between the \ce{CO} and \ce{CH4} icelines. Even in this region, the timescale is $>3$ Myr.  

The timescales for these processes scale inversely with the
cosmic-ray or X-ray ionisation rate. As such, in regions with lower
ionisation rates \ce{CO} destruction can be slower than described
here. The reverse is also true, so if locally produced energetic
particles play an significant role, the \ce{CO} destruction
timescales become shorter.

In conclusion, for a nominal cosmic-ray ionisation rate of $10^{-17}$
s$^{-1}$, CO can be significantly destroyed (orders of magnitude) in
the first few Myr of disk evolution under cold conditions. This
conclusion is robust against variations in chemical parameters, except
if the barrier width for the CO + H tunneling becomes too large. Under
warm conditions (T > 30 K), the CO abundance is lowered by a factor of a few at most.

\subsection{Implications for observations}

Mapping the results from the chemical models summarized above onto the physical structure seen in Fig.~\ref{fig:Model_over} results in Fig.~\ref{fig:COmapped}. It is clear
that \ce{CO} can be destroyed over a region of the disk that contains
a significant amount of mass.
For the T-Tauri disk a portion of this
\ce{CO} destruction will be in a region of the disk where {\ce{CO}} is not
traceable by observations, as the {\ce{CO}} is frozen out. However, there
is also a significant region up to 30 K within and above the CO snow
surface where CO is depleted and where the emission from the less
abundant {\ce{^{18}O}} and {\ce{^{17}O}} isotopologues originates
\citep{Miotello2014}. 
Thus, the processes studied here can explain, at least in part, the
observations of low \ce{CO} abundances within the ice-line in TW Hya
\citep{Zhang2017}.

If CO destruction would also be effective in disk regions that are
irradiated by moderate amounts $(< 1$ $G_0 )$ of (interstellar) UV
radiation, then a region responsible for $\sim 20$\% of the emission can be
depleted in CO by $\sim 1$ order of magnitude by 3 Myr, as seen in
Fig.~\ref{fig:COmapped}. The other $\sim 80$\% is also likely affected due to mixing of CO
poor gas from below the iceline into the emitting region (see Sec.~\ref{ssc:dynam}). This would
result in a drop of the \ce{C^{18}O} isotopologue flux between 20\% and an order
of magnitude depending on the vertical mixing efficiency. Rarer
isotopologues, whose flux comes from lower, colder layers, would be
more severely affected. The T-Tauri model shown in Figs.~\ref{fig:Model_over}~and~\ref{fig:COmapped} 
assume a total source luminosity of around $0.3 L_\odot$ representative
of the bulk of T-Tauri stars in low mass star-forming regions \citep{Alcala2017}. For disks around stars with a lower luminosity or with
higher ages, the effect of CO destruction would be even larger.

The warm upper layers exposed to modest UV have also been modelled by
\cite{Schwarz2018}. They find that the CO abundance is generally high
 in these intermediate layers, with significant CO destruction
only taking place for high X-ray or cosmic ray ionisation rates. This
is consistent with our Fig.~\ref{fig:COmapped} and Fig.~\ref{fig:DALI_full}.

For Herbig sources, only a small region of the disk falls within the
range of parameters studied here and an even smaller region is at
temperatures under 30 K, while also being shielded
(Fig.~\ref{fig:Model_over}). Because Herbig disks host more luminous
stars than their T-Tauri counterparts, they also tend to be
warmer. This means that a smaller mass fraction of the disk has
temperatures between 10 and 30 K, where \ce{CO} can be efficiently
destroyed. As such our chemical models predict that Herbig disk have a
\ce{CO} abundance that is close to canonical, consistent with
observations by \cite{Kama2016}.

Young disks around T Tauri stars are expected to be warmer due to the
accretional heating \citep{Harsono2015}. They are also younger than
the time needed to significantly convert \ce{CO} to other species with
grain surface reactions. As such young disks should close to canonical \ce{CO} abundances, as seems to be observed for at least some disks
\citep{vantHoff2018}.

\subsection{Observing chemical destruction of \ce{CO}}

In our models, CO is mostly processed into species that are frozen-out
in most of the disk, complicating detection of these products with
sub-mm lines. Several of the prominent carbon reservoirs, like CO$_2$,
CH$_4$ and C$_2$H$_6$, are symmetric molecules without a dipole
moment, so they do not have detectable sub-mm lines. Even CH$_3$OH,
which does have strong microwave transitions, is difficult to detect
since the processes that get methanol ice off the grains mostly
destroy the molecule \citep[e.g.,][]{Bertin2016,Walsh2016}. Thus,
observing the molecules in the solid state would be the best proof of
this chemical processing. Another way to indirectly observe the effects chemical processing of \ce{CO} would be to compare to cometary abundances. This will be discussed in Sec.~\ref{ssc:dynam}.

Ice does not emit strong mid-infrared bands. However, for some highly
inclined disks, the outer disk ice content can be probed with infrared
absorption against the strong mid-infrared continuum of the inner
  disk with the line of sight passing through the intermediate height
  disk layers \citep{Pontoppidan2005}. The processes discussed here
will leave a chemical imprint on the ice absorption spectra. The
destruction pathways of \ce{CO} preferentially create \ce{CO2},
\ce{CH3OH}, \ce{CH4} and \ce{H2O}.  \ce{CO2} reaches a peak abundance
of $\sim7 \times 10^{-5}$, seven times higher than the initial
conditions, which would be detectable. A large \ce{CO2} ice
reservoir would therefore immediately point at a conversion of \ce{CO} to
\ce{CO2}, either in the disk or {\it en route} to the disk
\citep{Drozdovskaya2016}. The formation of \ce{CH4} and \ce{CH3OH}
happens mostly in the coldest regions of the disk ($< 20$ K) as such
observations of large fractions ($>20\%$ with respect to \ce{H2O}) of
\ce{CH4} and \ce{CH3OH} in the ice would point at a disk chemistry
origin of \ce{CO} destruction. If large amounts of \ce{CH4} or
\ce{CH3OH} are found in infrared absorption spectra, this would
also imply that vertical mixing plays a role in the chemical
processing, as this is needed to move the \ce{CH4} and/or \ce{CH3OH}
ice hosting grains away from the cold midplane up to the layers that can
be probed with infrared observations. For \ce{H2O} the increase is
generally smaller than a factor of two, so that \ce{H2O} ice is not a
good tracer of \ce{CO} destruction.

If the gas and dust in the midplane is not static, but moves towards
the star due to the accretion flow or radial drift of dust grains,
then it is expected that the products of \ce{CO} destruction are
thermally desorbed in the inner disk
\citep[see][]{Booth2017,Bosman2018}. This would cause an observable
increase in abundance. Current infrared observations of gaseous
\ce{CO2} emission do not show a signal consistent with an \ce{CO2}
enriched inner disk, but the optical thickness of the 15 $\mu$m
\ce{CO2} lines in the surface layers might hinder the observation of
\ce{CO2} near the disk midplane \citep{Salyk2011,
  Pontoppidan2014,Walsh2015, Bosman2017, Bosman2018}. \cite{Gibb2013}
have observed gaseous \ce{CH4} in absorption in the near-infrared in
GV Tau N.  Their probed \ce{CH4} is rotationally hot (750 K) and is
thus likely situated within the inner 1 AU of the disk. They rule out
a large column of \ce{CH4} at lower temperatures (100 K). However,
this disk is still embedded so it is possibly too young to have
converted \ce{CO} into \ce{CH4} \citep{Carney2016}.  Gaseous \ce{CH4}
and \ce{CH3OH} have not been detected with {\it Spitzer}-IRS in disks,
but should be detectable with \textit{JWST}-MIRI. If lines from these
molecules are brighter than expected for inner disk chemistry, it
could point at a scenario in which \ce{CO} is converted to \ce{CH4}
and \ce{CH3OH} in the cold ($<20$ K) outer-midplane regions of the
disk, and that the reaction products are brought into the inner disk
via accretion or radial drift.

\subsection{Interactions with disk dynamics}
\label{ssc:dynam}
The chemical timescales needed for an order of magnitude decrease of the \ce{CO} dependence are close to 2 Myr. This timescale is significantly longer
than the turbulent mixing timescales, even in the outer disk. As such
it is not inconceivable that either vertical mixing or gas accretion
influences the abundance of \ce{CO}. For both the \ce{sCO + sH} and
the \ce{sCO + sOH} route, the rate scales with the abundance of
\ce{CO} in the ice: the higher the abundance, the faster the rate,
although the dependence can be sub-linear. This means that, if
vertical mixing replenishes the disk mid-plane \ce{CO} reservoir, the
destruction of \ce{CO} can happen at a higher rate for a longer
time. This converts \ce{CO} to other species not only in the regions
where the processes described in this paper are effective, but also in
the regions around it. For mixing to have a significant effect at 100 AU, a turbulent $\alpha$ of $10^{-4}$ or higher is needed \citep[see, e.g.][]{Ciesla2010, Semenov2011, Bosman2018}.

The freeze-out of \ce{CO} on grains that are large enough to have settled below the \ce{CO} snow surface can also lower the abundance of \ce{CO} in the disk atmosphere. This process is thought to happen for \ce{H2O} to explain the low observed abundances of \ce{H2O} in the outer disk \citep{Hogerheijde2011, Krijt2016Water, Du2017}. Using a toy model \cite{Kama2016} showed that the \ce{CO} abundance in the disk atmosphere can be lowered by 1-2 orders of magnitude in the disk lifetime. This process can work in concert with the chemical destruction of \ce{CO} to lower the \ce{CO} abundance in the outer disk. If large grains are locking up a significant fraction of the \ce{CO} near the midplane outside of the \ce{CO} iceline then this \ce{CO} would come off the grains near the \ce{CO} iceline, leading to an strong increase in the \ce{CO} abundance. \cite{Zhang2017} show however, that for TW Hya, this is not the case: \ce{CO} also has a low abundance of $\sim 3\times 10^{-6}$ within the \ce{CO} iceline. Further observations will have to show if this is the case for all disks with a low \ce{CO} abundance, or if TW Hya is the exception and this low \ce{CO} abundance is caused by its exceptional age of $\sim 8$ Myr \citep{Donaldson2016}.

If the destruction of \ce{CO} within the \ce{CO} snowline is a general
feature of protoplanetary disks, then a mechanism to stop any \ce{CO}
locked up in grains from being released in the gas phase will need to
be considered. This is hard to do without actually halting radial
drift completely, for the same reasons that it is hard to stop \ce{CO2} from
desorbing off the grain at its respective snowline
\citep{Bosman2018}. 

The fact that comets in the solar system do contain significant
  amounts of \ce{CO} (up to 30\% with respect to H$_2$O)
  \citep{Mumma2011,LeRoy2015} suggests that these comets were formed
  in a cold CO rich environment, possibly before the bulk of the
  \ce{CO} was converted into other molecules. In other disks, such
  CO-rich grains must then have been trapped outside the CO  iceline
  to prevent significant CO sublimating in the inner regions.

\section{Conclusions}
We performed a kinetic chemical modelling study of the destruction of
\ce{CO} under UV shielded, cold (< 40 K) and dense
($10^{6}$--$10^{12}$ cm$^{-3}$) conditions. Both grain-surface and
gas-phase routes to destroy \ce{CO} are considered and their
efficiencies and timescales evaluated using a gas-grain chemical
network. Furthermore we studied the effects of the assumed ice
diffusion speed (through the diffusion-to-binding energy ratio,
$f_\mathrm{diff}$) and the assumed \ce{H} and \ce{H2} tunnelling
efficiency (through the tunnelling barrier width, $a_\mathrm{tunnel}$)
on the evolution of the \ce{CO} abundance both in the gas-phase and on
the grain-surface.  Our findings can be summarised as follows:

\begin{itemize}
\item \ce{CO} destruction is linearly dependent on the assumed \ce{H2} ionisation rate by energetic particles (cosmic-rays, X-rays) over a large region of the considered physical parameter space. Only high enough cosmic-ray ionisation rates, $> 5\times 10^{-18}$ s$^{-1}$ can destroy CO on a $<$3 Myr timescale (Sec.~\ref{sssc:COdestr}, Fig.~\ref{fig:CR_CO}).

\item The chemical processing of \ce{CO} is most efficient at low
  temperatures. A relation between disk temperature and measured \ce{CO} abundance is expected. The coldest disks would have the lowest
  \ce{CO} abundances; in contrast, flaring disks around luminous Herbig stars
  should close to canonical CO abundances.

\item At low temperatures, hydrogenation of \ce{CO} is efficient when \ce{CO} is fully frozen-out, leading to a reduction of the total \ce{CO} abundance by $\sim$ 2 orders assuming $\zeta_{\ce{H2}} = 10^{-17}$ s$^{-1}$. This route is only weakly dependent on the temperature and density, as long as \ce{CO} is fully frozen-out (Sec.~\ref{ssc:phys_param}, Fig.~\ref{fig:dens_temp_chem}). 

\item At temperatures of 20--30 K, just above the desorption temperature of \ce{CO}, formation of \ce{CO2} from the reaction of \ce{CO} with \ce{OH} on the ice is efficient. The \ce{CO} abundance can be reduced by two orders of magnitude in 2--3 Myr for $\zeta_{\ce{H2}} = 10^{-17}$ s$^{-1}$. The formation of \ce{CO2} is more efficient at higher densities (Sec.~\ref{ssc:phys_param}, Fig.~\ref{fig:dens_temp_chem}). 

\item Gas-phase destruction of \ce{CO} by \ce{He+}, eventually leading to the formation \ce{CH4} and \ce{H2O}, only operates on timescales $> 5$ Myr for $\zeta_{\ce{H2}} = 10^{-17}$ s$^{-1}$. Furthermore, this pathway is only effective at low densities ($< 10^9$ cm$^{-3}$) and in a small range of temperatures (15 -- 25 K). As such, this pathway is not important in the context of protoplanetary disk midplanes (Sec.~\ref{ssc:phys_param}, Fig.~\ref{fig:dens_temp_chem}).

\item The assumed tunnelling barrier width ($a_\mathrm{tunnel}$) strongly influences the speed of \ce{CO} hydrogenation with efficient tunnelling leading to fast hydrogenation of \ce{CO}. The \ce{CO} destruction due to \ce{CO2} formation is only weakly dependent on the assumed chemical parameters. Only when $f_\mathrm{diff}$ is increased above 0.35 can this reaction be slowed down (Sec.~\ref{ssc:Chemical_param}, Fig.~\ref{fig:All_chemstudy}).  

\item \ce{CO2}, \ce{CH3OH} and, on a longer timescale, \ce{CH4} are all abundantly formed in the regions where \ce{CO} is destroyed. Observations of anomalously high abundances of \ce{CH4}, \ce{CH3OH} or \ce{CO2} either in infrared absorption spectroscopy towards edge-on systems, or in infrared emission from the inner disk, can help in distinguishing the chemical pathway responsible for \ce{CO} destruction.

\item Vertical mixing can bring gas from the warmer, \ce{CO} richer layers to the lower colder layers, where \ce{CO} can be converted into \ce{CH3OH} or \ce{CO2}. This would allow for the chemical processes to also lower the \ce{CO} abundance in the higher, warmer layers of the disk. 

\end{itemize}

In conclusion, chemical reprocessing of \ce{CO} can have a significant
impact on the measured disk masses if sufficient ionising radiation is
present in the cold ($< 30$ K) regions of the disk. Further modelling
of individual disks will have to show to which degree chemical
processes are important and where other physical processes will need
to be invoked.

\section*{Acknowledgements}
Astrochemistry in Leiden is supported by the European Union A-ERC grant 291141 CHEMPLAN, by the Netherlands Research School for Astronomy (NOVA), by a Royal Netherlands Academy of Arts and Sciences (KNAW) professor prize. CW acknowledges the University of Leeds for financial support.

\bibliographystyle{aa}
\bibliography{../../Literature/Lit_list}

\appendix

\section{Dali protoplanetary disk models}
\label{app:Dali_models}

For Fig.~\ref{fig:Model_over} two DALI models \citep{Bruderer2012,Bruderer2013} have been run to generate maps of the temperature and radiation field for the disks given the gas and dust density structures and stellar spectra as inputs. The input parameters can be found in Table~\ref{tab:Dali_param}. In the following the modelling procedure is shortly reiterated.

\begin{figure*}
\includegraphics[width =\hsize]{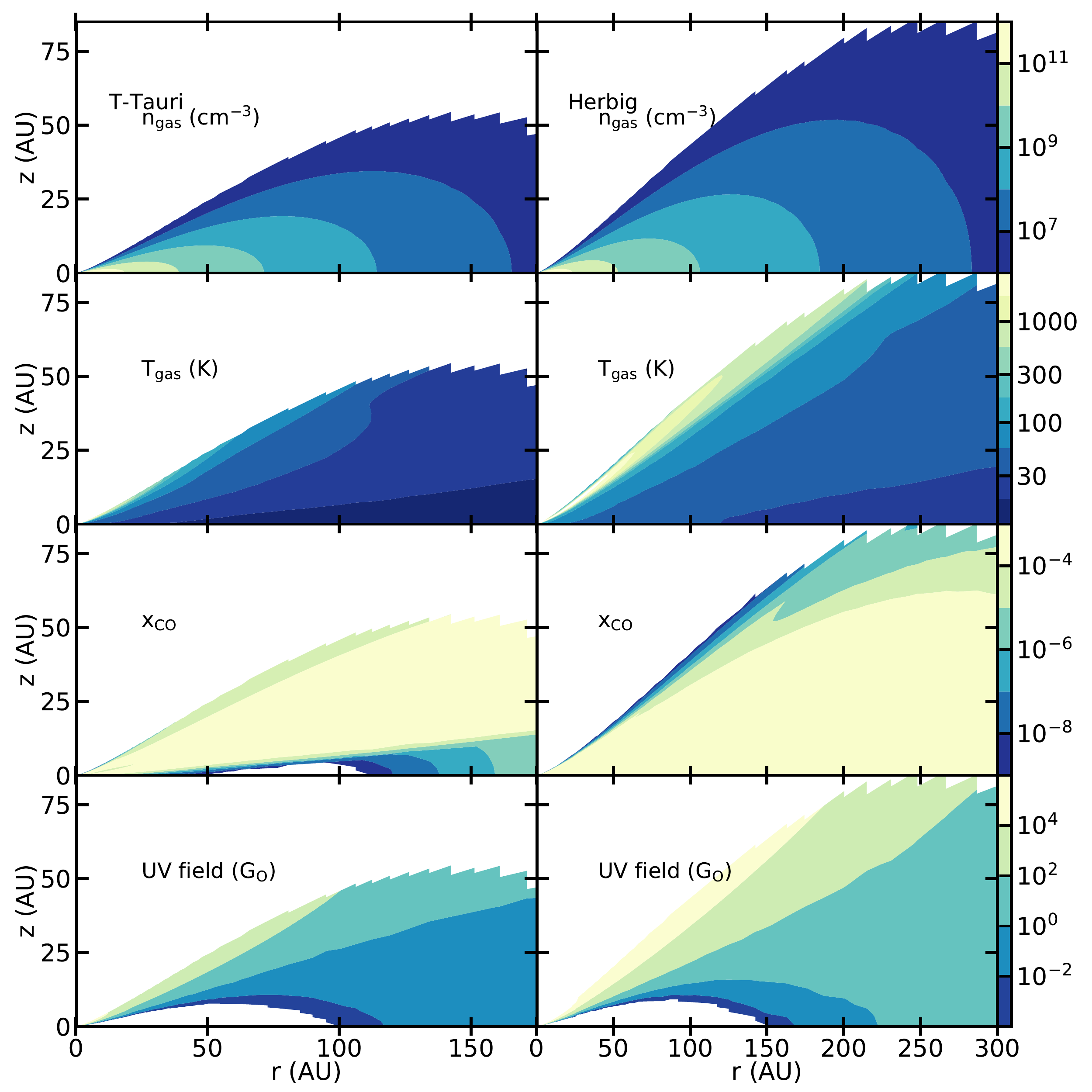}
\caption{\label{fig:DALI_full} Gas number density, gas temperature, gas-phase CO fractional abundance and UV radiation field for the T-Tauri and Herbig models. A full gas-grain model was not considered for the CO abundance. Freeze-out, desorption, gas-phase reactions and photo dissociation are included in the chemical computation. CO abundance after 1 Myr of chemical evolution is shown.}
\end{figure*}

Radially the gas and dust are distributed using a tapered powerlaw, 
\begin{equation}
\Sigma_{\mathrm{gas}} = \Sigma_c \left(\frac{R}{R_c})\right)^{-\gamma} \exp\left[-\left(\frac{R}{R_c}\right)^{2-\gamma}\right],
\end{equation}
\begin{equation}
\Sigma_{\mathrm{dust}} = \frac{\Sigma_c}{\mathrm{g/d}} \left(\frac{R}{R_c})\right)^{-\gamma} \exp\left[-\left(\frac{R}{R_c}\right)^{2-\gamma}\right].
\end{equation}
Vertically the gas and dust are distributed according to a Gaussian, the dust is divided into a "large" grain and a "small" grain population (see, Table~\ref{tab:Dali_param}, the large grain population contains a fraction $f_\mathrm{large}$ of the mass.
\begin{equation}
\rho_\mathrm{gas}(R,\Theta) = \frac{\Sigma_\mathrm{gas}(R)}{\sqrt{2\pi}Rh_\mathrm{gas}(R)}\exp\left[-\frac12\left(\frac{\pi/2 - \Theta}{h(R)}\right)^2\right],
\end{equation}
\begin{equation}
\rho_\mathrm{dust,\ small}(R,\Theta) = \frac{\left(1-f_\mathrm{large}\right)\Sigma_\mathrm{dust}(R)}{\sqrt{2\pi}Rh_\mathrm{gas}(R)}\exp\left[-\frac12\left(\frac{\pi/2 - \Theta}{h_\mathrm{gas}(R)}\right)^2\right],
\end{equation}
\begin{equation}
\rho_\mathrm{dust,\ large}(R,\Theta) = \frac{f_\mathrm{large}\Sigma_\mathrm{gas}(R)}{\sqrt{2\pi}Rh_\mathrm{large}(R)}\exp\left[-\frac12\left(\frac{\pi/2 - \Theta}{h_\mathrm{large}(R)}\right)^2\right],
\end{equation}
where $h_\mathrm{gas}(R) = h_c(R/R_c)^\psi$ and $h_\mathrm{large}(R) = \chi h_\mathrm{gas}(R)$. The smaller scale height for the larger grains mimics a degree of settling. The dust temperature and the radiation field are calculated using the continuum ray-tracing module of DALI. Fig.~\ref{fig:DALI_full} shows the density, dust temperature, radiation field and fractional abundance of CO for the DALI models. The CO abundance map does not include the effects of the grain surface chemistry discussed in this paper. 

\begin{table*}
\centering
\caption{\label{tab:Dali_param} Adopted model parameters for T-Tauri and Herbig disks.}
\begin{tabular}{l l c c }
\hline
\hline
Parameter &  & T-Tauri & Herbig\\
\hline
Star\\
Mass & $M_\star$ [$M_\odot$] & 0.2 & 2.5\\
Luminosity & $L_\star$ [$L_\odot$] & 0.3 & 20\\
Effective temperature & $T_\mathrm{eff}$ [K] & 3500 & 10000 \\
Accretion luminosity & $L_\mathrm{accr}$ [$L_\odot$] & 0.04 & 0 \\
Accretion temperature & $T_\mathrm{accr}$ [K] & 10000 & --\\
\hline
Disk\\
Disk Mass ($g/d = 100$) & $M_\mathrm{disk}$ [$M_\odot$]& 0.01 & 0.024\\
Surface density index & $\gamma$ &  1.0 &  1.0 \\
Characteristic radius&  $R_c$ [AU] & 35 &  75\\
Inner radius&  $R_\mathrm{in}$ [AU] & 0.07 & 0.221\\
Scale height index & $\psi$ & 0.3 & 0.3\\
Scale height angle & $h_c$ [rad]& 0.1 & 0.1\\
Large grain height scaling & $\chi$ & 0.2 & 0.2 \\
\hline
Dust properties\tablefootmark{a}\\
Large grain fraction & $f_\mathrm{large}$ & 0.90 & 0.90 \\ 
Size small & a[$\mu m$] & 0.005 -- 1 & 0.005 -- 1\\
Size large & a[$\mu m$] & 0.005 -- 1000 & 0.005 -- 1000\\
Size distribution & & $\mathrm{d}n/ \mathrm{d}a \propto a^{-3.5} $\\
Composition & & ISM \\
Total gas-to-dust ratio & $\mathrm{g/d}$&100 \\
\hline
\end{tabular}
\tablefoot{
\tablefoottext{a}{Dust composition is taken from \cite{Draine1984} and \cite{Weingartner2001}.}
}
\end{table*}

\section{Chemical model}
The abundance evolution is computed by a chemical solver based on its counterpart in the DALI code \citep{Bruderer2012,Bruderer2013}. To solve the set of differential equations, CVODE from the SUNDAILS suite is used \citep{Hindmarsh2005sundials}. CVODE was chosen over LIMEX, the which normally used in DALI, as CVODE is faster and more stable on the very stiff grain-surface chemistry models as well as being thread safe, enabling the calculation of multiple chemical models in parallel with OpenMP\footnote{\url{http://www.openmp.org/}}.

\subsection{Initial abundances}
\label{App:init_abu}
Table~\ref{tab:initabu} shows the input volatile abundances used in our chemical model. These abundances are depleted in \ce{Si} and \ce{S} but not in \ce{Fe} with respect to solar. \ce{Fe} will be frozen-out and will not play an active part in the chemistry. 
\begin{table}
\centering
\caption{\label{tab:initabu} Initial gas-phase abundances for the chemical network}
\begin{tabular}{l c | l c}
\hline
\hline
molecule & abundance & molecule & abundance \\
\hline
\ce{H2} & 0.5      & \ce{He} &  9.75(-2) \\
\ce{NH3} & 1.45(-6) & \ce{H2O} & 1.18(-4) \\
\ce{CO} & 1(-4) & \ce{N2} & 2(-5) \\
\ce{Si} & 4(-10) & \ce{CH3OH} & 1(-6) \\
\ce{H2S} & 1.91(-8) & \ce{CO2} & 1(-5)\\
\ce{Fe} & 4.27(-7) & grains & 2.2(-12)\\
\hline
\end{tabular}
\end{table}

Table~\ref{tab:Bindingenergies} enumerates all the binding energies used in the model. The list is sorted (left to right, top to bottom) according to molecular mass. Binding energies come from the recommended values from \cite{Penteado2017}  For \ce{NH}, \ce{NH2}, \ce{CH}, \ce{CH2} and \ce{CH3}, the binding energy was increased compared with the values from \cite{Penteado2017}. \ce{NH} and \ce{NH2} were calculated using the method of \cite{Garrod2006}, while for \ce{CH}, \ce{CH2} and \ce{CH3} the binding energy are calculated by linearly scaling between the values for \ce{C} and \ce{CH4} to the number of hydrogen atoms.

\begin{table*}
\centering
\caption{\label{tab:Bindingenergies} Binding energies for all the species in the chemical network. }
\begin{tabular}{l c | l c | l c | l c }
\hline
\hline
molec. & $E_\mathrm{bind}$ (K) & molec. & $E_\mathrm{bind}$ (K) & molec. & $E_\mathrm{bind}$ (K) & molec. & $E_\mathrm{bind}$ (K) \\
\hline
\ce{H} &     600  & \ce{H2} &     430  & \ce{He} &     100  & \ce{C} &     800 \\ 
\ce{CH} &     873  & \ce{CH2} &     945  & \ce{N} &     800  & \ce{CH3} &    1018 \\ 
\ce{NH} &    1577  & \ce{CH4} &    1090  & \ce{NH2} &    2354  & \ce{O} &     800 \\ 
\ce{NH3} &    3130  & \ce{OH} &    2850  & \ce{H2O} &    5770  & \ce{C2} &    1600 \\ 
\ce{C2H} &    2137  & \ce{C2H2} &    2587  & \ce{CN} &    1600  & \ce{C2H3} &    3037 \\ 
\ce{HCN} &    3610  & \ce{HNC} &    2050  & \ce{C2H4} &    3487  & \ce{CO} &     855 \\ 
\ce{H2CN} &    2400  & \ce{N2} &     790  & \ce{Si} &    2700  & \ce{C2H5} &    3937 \\ 
\ce{CH2NH} &    3428  & \ce{HCO} &    1600  & \ce{SiH} &    3150  & \ce{C2H6} &    2300 \\ 
\ce{H2CO} &    2050  & \ce{NO} &    1600  & \ce{SiH2} &    3600  & \ce{CH2OH} &    4330 \\ 
\ce{CH3O} &    2655  & \ce{HNO} &    2050  & \ce{SiH3} &    4050  & \ce{CH3OH} &    4930 \\ 
\ce{O2} &    1000  & \ce{S} &    1100  & \ce{SiH4} &    4500  & \ce{HS} &    1500 \\ 
\ce{O2H} &    3650  & \ce{H2O2} &    5700  & \ce{H2S} &    2743  & \ce{C3} &    2400 \\ 
\ce{C3H} &    2937  & \ce{C2N} &    2400  & \ce{C3H2} &    3387  & \ce{H2CCC} &    2110 \\ 
\ce{CH2CCH} &    3837  & \ce{HCCN} &    3780  & \ce{C2O} &    1950  & \ce{CH2CCH2} &    4287 \\ 
\ce{CH2CN} &    4230  & \ce{CH3CCH} &    4287  & \ce{SiC} &    3500  & \ce{CH3CN} &    4680 \\ 
\ce{HC2O} &    2400  & \ce{HCSi} &    1050  & \ce{CH2CO} &    2200  & \ce{CH3CHCH2} &    5187 \\ 
\ce{CNO} &    2400  & \ce{NH2CN} &    1200  & \ce{OCN} &    2400  & \ce{SiCH2} &    1100 \\ 
\ce{SiN} &    3500  & \ce{CH3CO} &    2320  & \ce{HCNO} &    2850  & \ce{HNCO} &    2850 \\ 
\ce{HNSi} &    1100  & \ce{HOCN} &    2850  & \ce{HONC} &    2850  & \ce{SiCH3} &    1150 \\ 
\ce{CH3CHO} &    3800  & \ce{CO2} &    2990  & \ce{CS} &    1900  & \ce{N2O} &    2400 \\ 
\ce{SiO} &    3500  & \ce{COOH} &    5120  & \ce{HCS} &    2350  & \ce{C2H5OH} &    5200 \\ 
\ce{CH3OCH3} &    3300  & \ce{H2CS} &    2700  & \ce{H2SiO} &    1200  & \ce{HCOOH} &    5000 \\ 
\ce{NO2} &    2400  & \ce{NS} &    1900  & \ce{C4} &    3200  & \ce{SO} &    2600 \\ 
\ce{C4H} &    3737  & \ce{C3N} &    3200  & \ce{C4H2} &    4187  & \ce{C4H3} &    4637 \\ 
\ce{HC3N} &    4580  & \ce{HNC3} &    4580  & \ce{C3O} &    2750  & \ce{CH2CHCCH} &    5087 \\ 
\ce{NCCN} &    1300  & \ce{SiC2} &    1300  & \ce{CH2CHCN} &    5480  & \ce{SiC2H} &    1350 \\ 
\ce{C2H4CN} &    5930  & \ce{C4H6} &    5987  & \ce{SiC2H2} &    1400  & \ce{SiNC} &    1350 \\ 
\ce{C2H5CN} &    6380  & \ce{C2S} &    5320  & \ce{Fe} &    4200  & \ce{CH3COCH3} &    3300 \\ 
\ce{CH2OHCO} &    6230  & \ce{COOCH3} &    3650  & \ce{C5} &    4000  & \ce{CH2OHCHO} &    6680 \\ 
\ce{CH3COOH} &    6300  & \ce{HCOOCH3} &    4000  & \ce{OCS} &    2888  & \ce{SiO2} &    4300 \\ 
\ce{SiS} &    3800  & \ce{C5H} &    4537  & \ce{C4N} &    4000  & \ce{C5H2} &    4987 \\ 
\ce{CH3C4H} &    5887  & \ce{S2} &    2200  & \ce{SO2} &    5330  & \ce{SiC3} &    1600 \\ 
\ce{CH3C3N} &    6480  & \ce{HS2} &    2650  & \ce{SiC3H} &    1650  & \ce{H2S2} &    3100 \\ 
\ce{C3S} &    3500  & \ce{C6} &    4800  & \ce{C6H} &    5337  & \ce{C5N} &    4800 \\ 
\ce{C6H2} &    5787  & \ce{HC5N} &    6180  & \ce{SiC4} &    1900  & \ce{C6H6} &    7587 \\ 
\ce{C4S} &    4300  & \ce{C7} &    5600  & \ce{C7H} &    6137  & \ce{C7H2} &    6587 \\ 
\ce{CH3C6H} &    7487  & \ce{CH3C5N} &    7880  & \ce{C8} &    6400  & \ce{C8H} &    6937 \\ 
\ce{C7N} &    6400  & \ce{C8H2} &    7387  & \ce{HC7N} &    7780  & \ce{C9} &    7200 \\ 
\ce{C9H} &    7737  & \ce{C9H2} &    8187  & \ce{CH3C7N} &    9480  & \ce{C10} &    8000 \\ 
\ce{C10H} &    8537  & \ce{C10H2} &    8987  & \ce{C9N} &    8000  & \ce{HC9N} &    9380 \\ 
\ce{C11} &    8800  & & & & & \\
\hline
\end{tabular}
\end{table*}
\subsection{{\ce{H2}} formation rate}
\label{app:H2form}
The formation of \ce{H2} is implemented following \cite{Cazaux2004}, who give the \ce{H2} formation rate as (in \ce{H2} molecules per unit volume per unit time):
\begin{equation}
R_{\ce{H2}} = \frac{1}{2} n_{\ce{H}} v_{\ce{H}} N_\mathrm{grain} \pi a_\mathrm{grain}^2 \epsilon_{\ce{H2}} S_{\ce{H}}(T),
\end{equation}
where $n_{\ce{H}}$ is the number density of gaseous \ce{H}, $v_{\ce{H}}$ is the thermal velocity of atomic hydrogen, $N_\mathrm{grain}$ is the absolute number density of grains,  $a_\mathrm{grain}$ is the grain radius. $S_{\ce{H}}(T)$ is the sticking efficiency \citep{Cuppen2010}, given by, 
\begin{equation}
S_{\ce{H}}(T) = \frac{1}{1.0+0.04\sqrt{T_\mathrm{gas}+T_\mathrm{dust}}+2\times 10^{-3} T_\mathrm{gas}+8\times 10^{-6}T_\mathrm{gas}^2}. 
\end{equation}
Furthermore, $\epsilon_{\ce{H2}}$ is the \ce{H2} recombination efficiency given by:
\begin{equation}
\begin{split}
\epsilon_{\ce{H2}} = \left(1 + \frac{1}{4} \left(1 + \sqrt{\frac{E_{\mathrm{H}_C} - E_S}{E_{\mathrm{H}_P} - E_S}}\right)^2 \exp\left[-\frac{E_S}{kT_\mathrm{dust}}\right] \right)^{-1} \\
\times \left[1 + \frac{v_{\mathrm{H}_C}}{2F}\exp\left(-\frac{1.5E_{\mathrm{H}_C}}{kT_\mathrm{dust}}\right)\left(1 + \sqrt{\frac{E_{\mathrm{H}_C} - E_S}{E_{\mathrm{H}_P} - E_S}}\right)^2\right]^{-1}, 
\end{split}
\end{equation}
where $E_{\mathrm{H}_P} = 600$ K, $E_{\mathrm{H}_C} = 10000$ K and $E_S = 200$ K are the energies of a physisorbed \ce{H} (\ce{H_P}), chemisorbed \ce{H} (\ce{H_C}) and the energy of the saddle point between the previous two, respectively. $F = 10^{-10}$ monolayers s$^{-1}$ is the accretion flux of \ce{H} on the ice for the \ce{H2} formation.

\subsection{Calculation of grain-surface rates}
\label{app:grainsurf}
To calculate the reaction rate coefficients of grain-surface reactions, the mobility of the molecules over the grain-surface needs to be known. This requires knowledge of the vibrational frequency of a molecule in its potential well on the grain-surface (assuming a harmonic oscillator. This is the frequency at which the molecule will attempt displacements:
\begin{equation}
\label{eq:vibfreq}
\nu_{\ce{X}} = \sqrt{ \frac{N_\mathrm{sites} E_\mathrm{X, bind}}{2\pi^3a_\mathrm{grain}^2 m_X}}, 
\end{equation}
where $N_\mathrm{sites}$ is the number of adsorption sites per grain, $E_\mathrm{X,\ bind}$ is the binding energy of the molecule to the grain, $a_\mathrm{grain} = 10^{-5}$ cm is the grain radius and $m_X$ is the mass of the molecule. The hopping rate is the vibrational frequency multiplied by the hopping probability, which in the thermal case is:
\begin{equation}
R_\mathrm{hop,\ X} = \nu_{\ce{X}} \exp\left(- \frac{f_{\mathrm{diff}} E_\mathrm{bind, X}}{kT}\right),
\end{equation}
where $k$ is the Boltzmann constant and $T$ is the dust-grain temperature. For atomic and molecular hydrogen, tunnelling is also allowed:
\begin{equation}
R_\mathrm{hop,\ tunnel,\ X} = \nu_{\ce{X}} \exp\left[-\frac{2 a_{\mathrm{tunnel}}}{\hbar}\sqrt{2\mu f_{\mathrm{diff}} E_\mathrm{bind, X}}\right].
\end{equation}
For atomic and molecular hydrogen, the fastest of these two rates is used which will almost always be the thermal rate. Only for models with a high $f_{\mathrm{diff}}$, low $a_{\mathrm{tunnel}}$ and low temperature is tunnelling faster than thermal hopping.

The rate coefficient for a grain-surface reaction between species X and Y forming product(s) Z on the grain-surface is given by:
\begin{equation}
\begin{split}
\label{eq:GS_full}
k(\ce{X + Y}, \ce{Z}) = \min\left[ 1, \left(\frac{N^2_\mathrm{act} N^2_\mathrm{sites} n_\mathrm{grain}}{n^2_\mathrm{ice}}\right)\right] P_\mathrm{reac}(\ce{X + Y}, \ce{Z}) \\
\left( \frac{R_\mathrm{hop,\ X}}{N_\mathrm{sites}} + \frac{R_\mathrm{hop,\ Y}}{N_\mathrm{sites}} \right) \\
  = C_\mathrm{grain} P_\mathrm{reac}(\ce{X + Y}, \ce{Z}) \left(R_\mathrm{hop,\ X} + R_\mathrm{hop,\ Y} \right) 
\end{split}
\end{equation}
where $N_\mathrm{act} = 2$ is the number of chemically active layers, $N_\mathrm{sites} = 10^6$ is the number of molecules per ice monolayer, $n_\mathrm{grain} = 2.2 \times 10^{-12} n_\mathrm{gas}$ is the number density of grains with respect to \ce{H2} and $n_\mathrm{ice}$ is the total number density of species on the grain.  
$P_\mathrm{reac}(\ce{X + Y}, \ce{Z})$ is the reaction probability for the reaction. This includes the branching ratios if multiple products are possible as well as a correction for reactions that have a barrier. 

The reaction probability is given by:
\begin{equation} 
P_\mathrm{reac}(\ce{X + Y}, \ce{Z}) = b_r(\ce{X + Y}, \ce{Z})
\exp\left(-\frac{E_\mathrm{bar}(\ce{X + Y}, \ce{Z})}{kT}\right),
\end{equation}
where $b_r(\ce{X + Y}, \ce{Z}))$ is the branching ratio, $E_\mathrm{bar}(\ce{X + Y}, \ce{Z})$ is the energy barrier for the reaction. If either \ce{H} or \ce{H2} is participating in the reaction, tunnelling is included:
\begin{equation} 
\begin{split}
P_\mathrm{reac,\ tun}(\ce{X + Y}, \ce{Z}) &= b_r(\ce{X + Y}, \ce{Z}) \\ 
&\exp\left(-\frac{2 a_{\mathrm{tunnel}}}{\hbar}\sqrt{2\mu E_\mathrm{bar}(\ce{X + Y}, \ce{Z})}\right).
\end{split}
\end{equation}
The largest of the tunnelling and thermal crossing probability is taken for the actual rate calculation, for most reactions tunnelling dominates in contrast with hopping, where thermal crossing dominates.

\subsection{Implications of modelling assumptions}
\label{app:modassump}

\subsubsection{Variations in the {\ce{H2}} formation rate}
The formation speed of \ce{H2} from atomic hydrogen is very important in our models as, together with the destruction rate of \ce{H2}, it sets the abundance of atomic hydrogen in the gas. For the formation rate, the prescription of \cite{Cazaux2004} is used. The formalism forces the atomic \ce{H} gaseous abundance to $\sim 1$ cm$^{-3}$ for $\zeta_{\ce{H2}} = 10^{-17}$ s$^{-1}$ irrespective of total gas density. A higher or lower atomic hydrogen abundance would strongly affect both the \ce{sCO + sH} route, which would increase in effectiveness with higher atomic hydrogen abundances, and the \ce{sCO + sOH} route, which would decrease in effectiveness with higher atomic hydrogen abundances. 

\subsubsection{Initial conditions}
The initial conditions for the chemistry are shown in Table.~\ref{tab:initabu}. The focus was on testing the \ce{CO} destruction, hence we start with \ce{CO} as the major volatile carbon reservoir, with trace amounts of \ce{CO2} (10\% of \ce{CO}) and \ce{CH3OH} (1\% of \ce{CO}). At the densities considered here, there is no strong chemical alteration before freeze-out and desorption are balanced. Furthermore, a small portion of \ce{N} is in \ce{NH3}, with the rest of the nitrogen in \ce{N2}.

The initial conditions can have a significant impact on the evolution of the \ce{CO} abundances. The amount of \ce{H2O} ice on the grains determines the formation rate of \ce{OH} on the grain, which in turn impacts the transformation of \ce{CO} to \ce{CO2}. If the chemical evolution is started with a larger portion of carbon in hydrocarbons than assumed in this work, with the excess oxygen put into \ce{H2O}, then a shorter \ce{CO} destruction time-scale will be found in the regions where \ce{CO2} formation is effective.

The time-scale of the \ce{CO + He+} route is especially sensitive to the initial abundances. Starting with a significant fraction of the \ce{CO} already incorporated into \ce{CO2}, such as predicted by \cite{Drozdovskaya2016} and as observed in ices in collapsing envelopes \citep{Pontoppidan2008}, can shorten chemical timescales significantly. Chemical timescales can be further shortened by the removal of \ce{N2} from the gas-phase, such as by starting the chemical model with \ce{NH3} as the dominant nitrogen carrier. 

The time-scale of the \ce{sCO + sH} route is not very sensitive to the initial abundances, as long as they are molecular. The maximal amount of \ce{CO} removal is a function of the amount of \ce{CO2} in the ice initially, but as hydrogenation of CO is at least two orders of magnitude faster than the dissociation of \ce{CO2} due to cosmic-rays, this means that even when \ce{CO2} is the dominant carbon reservoir, the \ce{CO} abundance can still be reduced below $10^{-6}$ by \ce{sCO + sH} even with the \ce{CO} replenishment.

\subsubsection{Chemically active surface}
All our models assume that only the molecules in the top two layers of our ices are participating in chemical reactions. The composition of these layers is assumed to be the same of that of the bulk ice. If the number of layers that can participate in chemical reactions is doubled, then the timescales for all grain-surface reactions decrease by a factor of four (see Equation~\ref{eq:GS_full}). However, most of our grain-surface reactions are not limited by the speed of the grain-surface reactions, but by the availability of H-atoms and OH-radicals in the ice. The availability of H-atoms is set by the arrival rate of \ce{H}-atoms on the grains, and this linearly depends on the total grain-surface but not on the number of chemically active layers. The availability of the \ce{OH} radical is primarily set by the destruction rate of \ce{H2O} which is independent of the ice thickness. As such the effect of changing the number of active layers is mostly in setting the exact location of the icelines as a larger number of active layers results in a faster sublimation rate at equal temperatures. Varying the number of active layers between one and eight changes the ice-line temperature by less than 1.5 K. This effect is similar to the effect of an one order of magnitude change in the total density. 

The total amount of grain-surface, and thus the assummed size of the grains does have some effect on the timescales of \ce{CO} destruction. The atomic hydrogen arrival rate on the grain scales with the total available surface, which decreases if the grain size is increased and the gas-to-dust mass ratio is held fixed. However, the \ce{H2} formation rate as calculated by \cite{Cazaux2004} also depends on the total surface area, thus if the total surface area is decreased, that means that the abundance of atomic hydrogen should increase, cancelling the effect of the grain size on the hydrogenation reaction of \ce{CO}. The production of \ce{CO2} actually increases with increasing grain size (and thus decreasing surface area). As the total grain-surface area decreases, a smaller amount of the cosmic-ray induced UV photons are absorbed by the dust grains, as such the dissociation rate for \ce{H2O} should be slightly higher, creating more \ce{OH} radicals and thus increasing the rate of \ce{CO2} production. 

\subsubsection{Mixed ice model}
The code assumes a perfectly mixed ice, that is to say that all species are assumed to be present in the upper chemically active layers at the relative abundances that they are included in the bulk of the ice. Observations of dark clouds indicate that in that stage ice mantles are not perfectly mixed, but have a layered structure \citep{Tielens1991,Boogert2015} with a CO-rich layer on top of the \ce{H2O}-rich layers. It is expected that ice mantles in protoplanetary disks are also layered. The perfect mixing assumption can have some profound effects on the chemistry of \ce{CO}. A layered ice below the \ce{CO} freeze-out temperature would have a higher fraction of \ce{CO} in the surface layers than currently assumed. This would initially speed-up the formation of \ce{CH3OH}, however, as the \ce{CO} in the top layers is converted into \ce{CH3OH}, the rate of \ce{CO} destruction would go down. At that point the \ce{CO} destruction rate would be dependent on the \ce{CO} exchange rate of the bulk ice and the surface layers which is expected to be slower than the surface diffusion speed. Just above the \ce{CO} freeze-out temperature the top layers of the ice would be made-up of mostly \ce{CH3OH}, \ce{CO2} and \ce{CH4}. Since the top layers have very little \ce{H2O}, this would lead to very low abundances of \ce{OH} radicals in the surface layers and might thus quench the \ce{sCO +sOH} route to form \ce{CO2}.

Mixing of an ice through bulk diffusion is a slow process. However, in the high density midplanes another process can mix ices: grain-grain collisions. In large parts of the disk, the grain size distribution is in coagulation-fragmentation equilibrium. Especially the destructive collisions leading to cratering and fragmentation should be able to mix up the layered structure. 

Multi-phase models that include layering explicitly have been developed \citep[e.g.][]{Taquet2012,Furuya2016}. \cite{Drozdovskaya2016} compared the results from their two-phase (gas and ice) chemical model to the three-phase (gas, surface ice and bulk ice) chemical model from \cite{Furuya2015}. They conclude that the two-phase model and the three-phase model that includes swapping between bulk and surface ice produces comparable amounts of \ce{CO2} in the ice, but that if swapping is excluded, \ce{CO2} production is efficiently quenched. More detailed models of disk midplanes that included layering should be done in the future to better quantify the effects of assuming a fully mixed ice on the results presented here. 

\subsubsection{CO binding energy}

The binding energy of \ce{CO} depends strongly on the surface on which it is frozen \citep{Collings2003,Acharyya2007}. Here the binding energy of \ce{CO} on a pure \ce{CO} ice surface is used, which is applicable to the cold ($< 20$ K) regions, where CO is primarily in the ice. In warmer regions however, \ce{CO} will mostly be bound to \ce{CO2}, \ce{CH3OH} and \ce{H2O}, slightly increasing the binding energy of \ce{CO} \citep{Collings2003,Noble2012,Acharyya2007}. The higher binding energy, and thus slower \ce{CO} diffusion, moves upwards and broadens the temperature range over which the reaction to form \ce{CO2} (Equation~\ref{eq:CO2form}) is efficient. 

\end{document}